%% file: main.tex
\documentclass{article}

\usepackage{arxiv}

\usepackage[utf8]{inputenc} 
\usepackage[T1]{fontenc}    
\usepackage{url}            
\usepackage{hyperref}       
\usepackage{array}
\usepackage{booktabs}       
\usepackage{amsfonts}       
\usepackage{nicefrac}       
\usepackage{microtype}      
\usepackage{graphicx}
\usepackage{natbib}
\usepackage{doi}
\usepackage{float}
\usepackage{amsmath}
\usepackage{bbm}
\usepackage{cleveref}
\crefname{section}{Section}{Sections}
\Crefname{section}{Section}{Sections}
\crefname{figure}{Figure}{Figures}
\Crefname{figure}{Figure}{Figures}
\crefname{table}{Table}{Tables}
\Crefname{table}{Table}{Tables}
\crefname{equation}{Eq.}{Eqs.}
\Crefname{equation}{Equation}{Equations}

\usepackage{amssymb}
\usepackage{caption, subcaption}
\usepackage[bottom]{footmisc} 
\usepackage{xurl} 
\usepackage{lscape} 
\usepackage[section]{placeins} 
\usepackage{afterpage} 
\usepackage[table]{xcolor} 
\usepackage{threeparttable}
\usepackage{multirow}
\title{What Do Bitcoin Premiums Measure? Evidence from Global P2P Markets}

\author{ Yanan Niu\\
	EPFL\\
	\texttt{yanan.niu@epfl.ch} \\
}

\renewcommand{\headeright}{\textit{arXiv}}
\renewcommand{\shorttitle}{What Do Bitcoin Premiums measure?}

\hypersetup{
pdftitle={What Do Bitcoin Premiums measure?},
pdfauthor={Yanan Niu},
}

\begin{document}

\maketitle

\begin{abstract}
This paper studies what Bitcoin (BTC) premiums in peer-to-peer (P2P) markets measure. Using transaction-level data from LocalBitcoins, we construct BTC premiums for 80 currencies relative to the U.S. dollar and relate them to blockchain transaction conditions, centralized crypto market (CEX) conditions, cross-border payment frictions, and foreign exchange (FX) markets.

We show that these premiums reflect both trading frictions within crypto markets and local frictions in access to cross-border payments. They vary systematically with blockchain conditions and broader crypto market conditions, including BTC returns and volatility, and they are larger in countries facing greater frictions in conventional cross-border payment channels. This pattern is especially pronounced in economies with binding institutional constraints, i.e., tight capital controls and non-floating exchange-rate regimes, consistent with greater reliance on P2P crypto markets as an alternative cross-border payment channel.

We further show that rising FX pressure is absorbed mainly through prices rather than trading volumes, and that P2P BTC premiums predict subsequent official exchange rate depreciation. Although premium levels differ across countries, their predictive content remains broadly similar. 

Overall, P2P BTC premiums reflect limits to arbitrage across crypto trading venues, especially where formal cross-border payment channels are more constrained, and they also embed forward-looking information about currency depreciation.
\end{abstract}

\keywords{Bitcoin premiums \and Peer-to-peer crypto markets \and Exchange rates \and Capital controls \and Cross-border payments \and Market segmentation}
\input{Sections/Introduction}
\input{Sections/background}
\input{Sections/Empirical_framework}
\input{Sections/Data}

\input{Sections/Result_2026}
\input{Sections/Price_discovery}
\input{Sections/Conclusion}

\newpage

\bibliographystyle{unsrtnat}
\bibliography{Reference}  

\input{Sections/Appendix}
\end{document}

%% file: Sections/Introduction.tex
\section{Introduction}
Despite the global and digital nature of cryptocurrency markets, Bitcoin (BTC) often trades at persistently different prices across countries. Once local prices are converted to USD at prevailing spot exchange rates, they can diverge from U.S. exchange prices. A well-known example is the ``Kimchi premium'', where BTC on Korean exchanges (quoted in KRW and converted to USD) traded at a premium of over 50\% relative to U.S. exchanges in early 2018 \citep{Choi2018}. More broadly, \citet{Makarov2020} document large and recurrent price deviations across various exchanges and geographic regions.

These price differentials are not short-lived deviations driven by noise. Instead, they are highly persistent and economically significant. A growing literature attributes them to institutional frictions and cross-border constraints that limit arbitrage \citep{crepelliere2023}. In parallel, especially in emerging markets, economic agents increasingly rely on peer-to-peer (P2P) crypto platforms to circumvent capital controls and facilitate cross-border transactions \citep{Rogoff2023}.

These observations point to a broader question: \textbf{what do these persistent BTC price premiums measure?} Do they simply reflect limits to arbitrage, or do they also encode information about underlying macroeconomic conditions?

Existing studies document several determinants of cryptocurrency price deviations. BTC premiums tend to widen during periods of cryptocurrency appreciation and are correlated with capital controls \citep{Choi2018, Makarov2020}. Blockchain microstructure frictions, such as network congestion, can further amplify these deviations \citep{Choi2018}. However, it remains unclear whether these price differentials primarily reflect frictions within crypto markets, frictions in cross-border payments, or exchange rate volatility. More broadly, the literature has not established whether BTC premiums contain information about future currency movements.

Another limitation of the existing literature is its focus on centralized exchanges (CEXs), leaving the behavior of price deviations in peer-to-peer (P2P) markets largely unexplored. P2P markets are especially relevant in settings where access to formal cross-border channels is restricted and may therefore better capture local frictions that CEX-based studies overlook.

In addition, prior empirical designs often face a trade-off between clean identification and sample coverage. Studies of arbitrage mechanisms often exclude illiquid or capital-restricted markets, while those focusing on capital controls often rely on a limited set of extreme cases. As a result, the cross-country variation needed to quantify the effects of institutional frictions and their interaction with FX conditions remains underexplored.

In this paper, we show that persistent BTC premiums in P2P markets reflect more than limits to arbitrage: they also capture local frictions in cross-border payments and exchange rate pressure. Using transaction-level data from LocalBitcoins, we construct a daily panel of P2P BTC premiums across 80 currencies over 2017 -- 2023.  Starting from P2P prices and FX rates, we introduce a CEX benchmark that allows us to express premiums relative to BTC prices on global CEXs. We then decompose each premium into two components: a global component, defined as the USD-denominated spread between P2P and CEX prices, and a local component, defined as currency-specific deviations from that global spread.

By linking P2P markets to both CEX and FX markets, we are able to separately identify the roles of blockchain transaction frictions, global crypto market conditions, local cross-border payment frictions, and FX conditions in shaping P2P pricing. Our empirical analysis yields three main findings:

\textbf{First, P2P premiums respond more strongly to FX volatility and formal cross-border payment frictions in economies with tight capital controls and non-floating exchange-rate regimes.} This pattern is consistent with P2P markets functioning as an alternative channel for capital flows when traditional financial systems are constrained.

\textbf{Second, when FX volatility rises, P2P market adjustment occurs primarily through prices rather than trading volumes.}  In normal times, P2P markets adjust through joint price and quantity responses. However, when exchange rate volatility rises, this adjustment mechanism weakens: trading volumes become relatively inelastic while premiums widen sharply, consistent with limited supply responsiveness in local markets.

\textbf{Third, BTC premiums contain forward-looking information about exchange rate movements.} Despite substantial cross-country variation in levels, premiums are positively correlated with future depreciations in official exchange rates (OERs). This suggests that P2P markets incorporate information about currency pressure that is not fully reflected in current OERs.

The remainder of the paper is organized as follows. \Cref{sec:bg} reviews the background and related literature. \Cref{sec:framework} presents the conceptual framework and empirical strategy. \Cref{sec:data} describes the data and variable construction, and presents stylized facts with selected country case studies. \Cref{sec:results} reports the main empirical results and heterogeneity analyses across institutional environments, while \cref{sec:var} examines the dynamic relationship between P2P premiums and OERs using VAR models. Finally, \cref{sec:Conclusion} concludes.

%% file: Sections/background.tex
\section{Institutional Background and Related Literature}\label{sec:bg}
\subsection{The P2P Cryptocurrency Market}
Compared to CEXs that rely on automated limit order books, P2P platforms facilitate cryptocurrency transactions in a fundamentally decentralized manner. We focus on LocalBitcoins.com (LB), historically one of the most prominent global P2P BTC marketplaces. On LB, users post customized advertisements specifying prices, volumes, and acceptable payment methods to buy or sell BTC. Trades are executed bilaterally once a counterparty accepts an offer. LB's low entry barriers, a moderate 1\% trading fee, and support for a broad spectrum of alternative payment channels---including digital wallets, domestic bank transfers, and cash\footnote{Cash-settled trades were discontinued in June 2019 due to stricter regulations on money laundering. See \url{https://twitter.com/LocalBitcoins/status/1135872083962081281}, last accessed in Mar. 2026.}---made it highly accessible, especially for agents seeking to bypass conventional banking bottlenecks. 

Crucially, LB supported over 100 fiat currencies, significantly outnumbering most CEXs. This extensive coverage organically reflects underlying local market conditions, making the P2P platform an ideal empirical setting for evaluating cross-country institutional frictions.

\subsection{Related Literature}
Our research primarily connects to three strands of literature. First, we build on the extensive body of work documenting violations of the law of one price and the extraction of implied exchange rates from cross-border assets. Historically, persistent price deviations have been observed in ``Siamese-twin'' cross-listed stocks \citep{Froot1999Debora} and closed-end country funds \citep{Bodurtha1995}. More directly related to our context, episodes of severe capital controls in emerging markets (e.g., Argentina, Malaysia, and Venezuela) have demonstrated that American Depositary Receipts\footnote{It is a type of stock listed in the United States but represents a specified number of shares in a foreign corporation.} can generate a shadow exchange premium that reflects underlying macroeconomic distress \citep{Eichler2009}. In this paper, we argue that cryptocurrencies, as globally homogeneous digital assets, provide a modern, high-frequency analogue to these traditional securities, allowing for a cleaner measurement of cross-country price differentials, with fewer confounding effects from corporate governance or taxation.

Second, our paper contributes to the literature on limits to arbitrage in segmented markets. Traditional finance studies emphasize that currency risks, liquidity constraints, and information barriers prevent arbitrageurs from closing cross-border price gaps \citep{Rosenthal1990, Froot1999Debora, Jong2009, Gagnon2010}. In the cryptocurrency domain, \citet{Choi2018} and \citet{Makarov2020} document that capital controls and blockchain microstructure frictions (e.g., network congestion and miner fees) amplify BTC premiums. However, by predominantly focusing on CEXs and arbitrage efficiency, this literature has paid limited attention to pricing dynamics within decentralized P2P platforms. We fill this gap by systematically evaluating how institutional constraints interact with exchange rate volatility and cross-border transaction costs to shape local P2P premiums.

Finally, our work engages with studies examining cryptocurrencies as conduits for capital flight and alternative financial intermediation. Existing empirical work has documented BTC's role in facilitating capital flight during periods of policy uncertainty or strict capital controls \citep{Ju2016, Hu2020}. More recently, \citet{Rogoff2023} estimate that approximately 11\% of trades on P2P platforms such as LocalBitcoins correspond to ``vehicle transactions'' associated with cross-border capital flows. We build on this literature by linking crypto markets to macroeconomic fundamentals. Rather than treating P2P premiums as static outcomes of capital flight, we show that they contain forward-looking information about currency pressure and exchange rate dynamics.

%% file: Sections/Empirical_framework.tex
\section{Conceptual Framework and Empirical Specification\label{sec:framework}}
\subsection{The P2P Premium and Cross-Market Arbitrage\label{sec:prem_def}}
\begin{figure}[t]
    \begin{subfigure}{.5\linewidth}
      \centering
      \includegraphics[width=.6\linewidth]{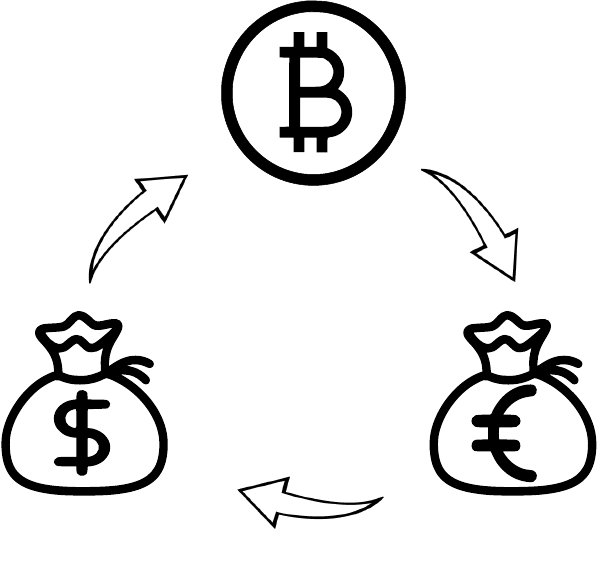}
      \caption{BTC trade flows under arbitrage}
      \label{fig:BTC_trade}
    \end{subfigure}
    \begin{subfigure}{.5\linewidth}
      \centering
      \includegraphics[width=.6\linewidth]{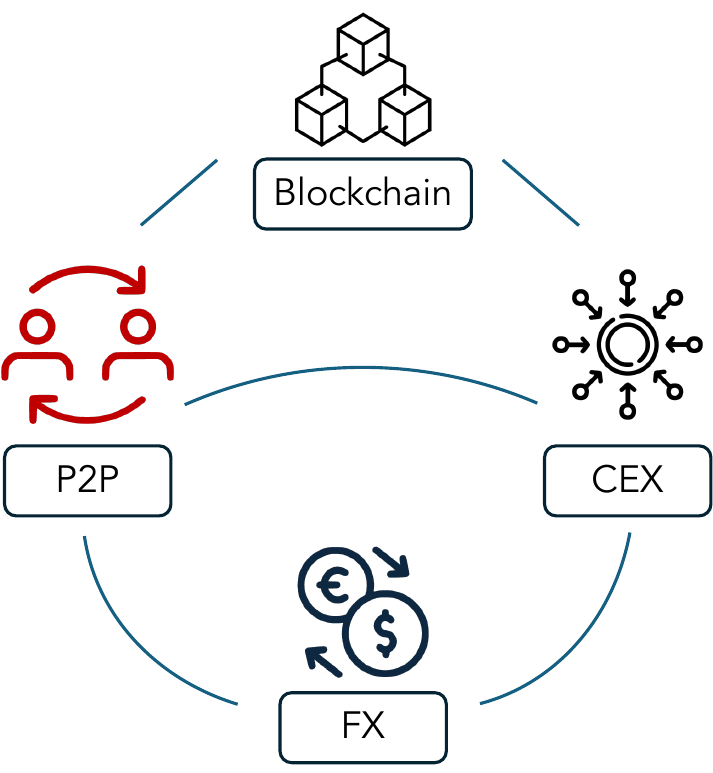}
      \caption{Market structure: blockchain, CEXs, P2P platforms, and FX markets}
      \label{fig:BTC_markets}
    \end{subfigure}
\caption{Bitcoin arbitrage across markets and trading venues}
\end{figure}
We define the P2P BTC premium as a cross-market price deviation. Let $P^{\mathrm{P2P}}_c$ denote the price of one BTC quoted in local currency $c$ on the P2P platform (LB), and let $P^{\mathrm{P2P}}_{\mathrm{USD}}$ denote the USD price of one BTC on the same platform. Their ratio defines the shadow exchange rate (SER), $\mathrm{SER}_c$. The P2P BTC premium, $\mathrm{Prem}^{\mathrm{P2P}}_c$, is the deviation of the SER from the OER:
\begin{align}
\mathrm{Prem}^{\mathrm{P2P}}_c &= \frac{\mathrm{SER}_c}{\mathrm{OER}_c} \nonumber \\
&= \frac{P^{\mathrm{P2P}}_c}{P^{\mathrm{P2P}}_{\mathrm{USD}}} \times \frac{1}{\mathrm{OER}_c}.
\label{eq:premium_def}
\end{align}

Under no-arbitrage conditions, $\mathrm{Prem}^{\mathrm{P2P}}_c = 1$. Any positive deviation from this parity gives rise to an arbitrage opportunity, which can be eliminated through a round-trip strategy across currencies. (\cref{fig:BTC_trade}): for example, an arbitrageur buys BTC in the cheaper USD market, sells it in the local-currency market at a premium, and converts the proceeds back to USD at the OER.

In practice, limits to arbitrage allow the premium to persist. On the currency-conversion side, capital controls, cross-border transaction costs, and exchange-rate uncertainty can prevent arbitrageurs from converting local currency back to USD at the OER. On the blockchain side, confirmation delays and transfer fees introduce execution risk. Together, these frictions sustain $\mathrm{Prem}^{\mathrm{P2P}}_c \neq 1$ and let it carry information about local economic conditions.

To benchmark P2P prices against a deeper, more liquid venue, we introduce the USD BTC price on CEXs, $P^{\mathrm{CEX}}_{\mathrm{USD}}$ (see \cref{fig:BTC_markets}). Using this benchmark, we rewrite \cref{eq:premium_def} to decompose the premium:

\begin{align}
\mathrm{Prem}^{\mathrm{P2P}}_c &=\frac{P^{\mathrm{P2P}}_c}{P^{\mathrm{P2P}}_{\mathrm{USD}}}\times \frac{P^{\mathrm{CEX}}_{\mathrm{USD}}}{P^{\mathrm{CEX}}_{\mathrm{USD}}} \times \frac{1}{\mathrm{OER}_c} \nonumber \\
    &= (\underbrace{\frac{P^{\mathrm{P2P}}_{\mathrm{USD}}}{P^{\mathrm{CEX}}_{\mathrm{USD}}}}_{\text{Baseline P2P Friction}})^{-1}\times \underbrace{\frac{P^{\mathrm{P2P}}_c/P^{\mathrm{CEX}}_{\mathrm{USD}}}{\mathrm{OER}_c}}_{\text{Currency-Specific Spread}}
    \label{eq:framework}
\end{align}
This decomposition separates the observed P2P premium into two distinct economic components:
\begin{itemize}
    \item \textbf{Baseline P2P Friction}: the gap between the USD P2P price and the USD CEX price. It reflects the liquidity and operational cost of trading on the P2P platform relative to CEXs.
    \item \textbf{Currency-Specific Spread}: the deviation of the local-currency P2P price from the USD CEX price after OER conversion. It captures country-specific distortions on the P2P platform.
\end{itemize}

Taking the logarithm of \cref{eq:framework} yields an additive structure:
\begin{equation}
\ln(\mathrm{Prem}^{\mathrm{P2P}}_c) = -\ln(\text{Baseline P2P Friction}) + \ln(\text{Currency-Specific Spread})
    \label{eq:framework_log}
\end{equation}
The log premium is additive in the two components, so each can be estimated in a linear panel specification below.

\subsection{Empirical Specifications}
We estimate the specifications below on a daily panel of 80 currencies traded on LocalBitcoins over 2017 -- 2023 (see \cref{sec:data}). Unless otherwise specified, all regressions are conducted at the daily frequency. For selected specifications (e.g., \cref{eq:hetero_trans}), we use weekly frequency for reasons discussed in the corresponding sections.

\subsubsection{Baseline Panel Model and Structural Decomposition}
Motivated by the structural decomposition in \cref{eq:framework_log}, we estimate the baseline specification three times---once for the total log P2P premium and once for each of its two components. This separates channels that act on the USD P2P--CEX gap (common across local markets) from channels that act on the local-currency spread (country-specific).

We study the determinants of these price deviations using the following dynamic panel specification:
\begin{equation}
    y_{c,t} = \rho \ln(\mathrm{Prem}^{\mathrm{P2P}}_{c,t-1}) + \boldsymbol{\beta}'\mathbf{X}_{c,t} + \boldsymbol{\gamma}'\mathbf{Z}_{t} + \mu_c + \tau_{w(t)} + \varepsilon_{c,t}
    \label{eq:prem_base}
\end{equation}
where the dependent variable $y_{c,t}$ is alternately set equal to: the log total premium $\ln(\mathrm{Prem}^{\mathrm{P2P}}_{c,t})$, the negative log baseline P2P friction $-\ln(\text{Baseline P2P friction}_{c,t})$, and the log currency-specific spread $\ln(\text{Currency-specific spread}_{c,t})$. Thus, \cref{eq:prem_base} represents three regressions with the same right-hand-side variables.

Because the dependent variables satisfy the accounting identity in \cref{eq:framework_log} and the three regressions share the same set of regressors, the OLS coefficients on the two components, $-\ln(\text{Baseline P2P friction}_{c,t})$ and $\ln(\text{Currency-specific spread}_{c,t})$, add up exactly to the coefficient on the total premium ($\ln(\mathrm{Prem}^{\mathrm{P2P}}_{c,t})$). To preserve this identity, we use the lagged total premium $\ln(\mathrm{Prem}^{\mathrm{P2P}}_{c,t-1})$ in all three regressions rather than the lags of the sub-components. As a result, the decomposition holds coefficient-by-coefficient: for each right-hand-side variable $k$, $ \hat{\delta}^{\text{total}}_k = \hat{\delta}^{\text{baseline}}_k + \hat{\delta}^{\text{specific}}_k.$ This allows us to decompose the response of the total premium to each regressor into contributions from the baseline P2P spread and the currency-specific component. 

$\mathbf{X}_{c,t}$ collects currency-$c$ time-varying controls: local P2P trading volume (platform liquidity) and log OER volatility (currency movements). $\mathbf{Z}_{t}$ collects day-$t$ global factors: CEX BTC return, realised CEX BTC volatility, and blockchain conditions (median confirmation time, transaction fees, on-chain volume). $\mu_c$ captures currency fixed effect and $\tau_{w(t)}$ denotes week fixed effects, where $w(t)$ maps time $t$ into calendar weeks. We use week rather than day effects because the global factors $\mathbf{Z}_{t}$ vary only over time: day fixed effects would absorb them entirely. Week effects soak up unobserved weekly-or-slower common movements (e.g.\ periods of elevated global risk aversion) while leaving within-week daily variation in $\mathbf{Z}_{t}$ to identify $\boldsymbol{\gamma}$.

To study whether price movements are accompanied by quantity adjustment, we estimate the same specification with log P2P trading volume as the dependent variable. Comparing the volume coefficients with the price coefficients allows us to assess whether a given regressor affects market adjustment primarily through quantities or through prices.

\subsubsection{Heterogeneous Transmission: Institutional and Transactional Frictions}
We next ask whether the pass-through from global factors to the local premium depends on the country's institutional environment and on the cost of the traditional cross-border alternative. We extend \cref{eq:prem_base} by interacting global factors with \textit{institutional} frictions (dummies for capital controls and exchange rate regimes) and \textit{transaction} costs (the cost of cross-border remittances).

We investigate the joint amplification effect of capital controls ($\mathrm{CC}$) and exchange rate regimes ($\mathrm{ERR}$) by augmenting \cref{eq:prem_base} with a triple interaction framework:
\begin{equation}
    y_{c,t} = \dots + \theta\big(\text{FX Vol}_{c,t} \times \mathrm{CC}_{c,t} \times \mathrm{ERR}_{c,t}\big) + \boldsymbol{\phi}' \text{Controls}_{c,t} + \mu_c + \tau_{w(t)} + \varepsilon_{c,t}
    \label{eq:hetero_inst}
\end{equation}
where $\text{Controls}_{c,t}$ contains all corresponding lower-order terms (main effects and all two-way interactions), so that $\theta$ captures the incremental effect of the three-way interaction. The specification asks whether the pass-through of FX volatility to the P2P premium is stronger in currencies with both tight capital controls and non-floating exchange rate regimes. The construction of the $\mathrm{CC}_c$ and $\mathrm{ERR}_c$ variables is described in \cref{sec:data}.

To capture transaction costs, we introduce a weekly proxy for remittance costs, $\mathrm{RC}_{c,w}$, see \cref{sec:remittance_def,Appendix:transformation} for details. To capture institutional constraints with one variable, we define $\mathrm{Constrained}_{c,w} = \mathbbm{1}\{\mathrm{CC}_{c,w} = 1 \ \text{and} \ \mathrm{ERR}_{c,w} \neq \text{Float}\}$, an indicator equal to one for countries with both tight capital controls and a non-floating exchange rate regime.  We therefore conduct the analysis on a weekly basis using the following specification:
\begin{equation}
    y_{c,w} = \dots + \xi \big(\mathrm{RC}_{c,w} \times \mathrm{Constrained}_{c,w}\big) + \boldsymbol{\eta}' \text{Controls}_{c,w} + \mu_c + \tau_{m(w)} + \upsilon_{c,w}
    \label{eq:hetero_trans}
\end{equation}

As in \cref{eq:prem_base}, we use time fixed effects at a lower frequency than the panel frequency: month fixed effects $\tau_{m(w)}$ absorb shocks common to all currencies in a given month, where $m(w)$ assigns each week to its corresponding calendar month. $\text{Controls}_{c,w}$ contains the main effects of $\mathrm{RC}_{c,w}$ and $\mathrm{Constrained}_{c,w}$, as well as weekly aggregates of the baseline covariates from \cref{eq:prem_base} (see \cref{sec:data} for details).

The coefficient $\xi$ is of primary interest. It tests whether an increase in remittance costs has a larger effect on the local P2P premium in currencies with binding institutional constraints than in unconstrained ones. A positive coefficient is consistent with limited access to conventional cross-border channels, leading to greater reliance on alternative channels such as P2P crypto markets.

%% file: Sections/Data.tex
\section{Data and Summary Statistics \label{sec:data}}
\subsection{Data}
To systematically analyze BTC premiums in P2P markets, we construct an unbalanced daily panel dataset that combines transaction-level P2P data with BTC prices from CEX, as well as macroeconomic and blockchain network variables.

\input{RegressionTables/transactions}
Our primary dataset consists of tick-level transaction records extracted via the public API of LB. A sample of the raw transaction data is illustrated in~\cref{Tab:tab_transaction}. We focus on the period from January 1, 2017, to February 9, 2023 (the cessation of LB's operations).\footnote{LB suspended its BTC trading services on February 9, 2023, citing unfavorable market conditions and regulatory pressures. See \url{https://localbitcoins.com/service_closure/}, last accessed in Oct. 2024. Although the platform is no longer active, its comprehensive historical data remains a uniquely valuable setting for analyzing global P2P pricing dynamics.} In total, we obtained 40,767,585 executed trades covering 81 fiat currencies.\footnote{We explicitly exclude 59 additional niche currencies that have fewer than 1,000 transactions over the sample period or lack reliable official exchange rate data.} We aggregate the daily currency-specific price of BTC, $P^{\text{P2P}}_{c}$, and the daily trading volume in BTC for each currency.

To capture institutional constraints and global crypto market conditions, we supplement the P2P data with the following datasets, all aligned to the daily frequency:
\begin{itemize}
    \item \textbf{Official Exchange Rates:} Daily FX spot rates obtained from the Thomson Reuters Refinitiv API. 
    \item \textbf{Blockchain Network Metrics:} Median confirmation time (in minutes), on-chain cost per transaction (in USD), and the number of confirmed transactions, collected from Blockchain.com.\footnote{See \url{https://www.blockchain.com/explorer/charts}, last accessed in Oct. 2024.} The first two metrics proxy for blockchain network efficiency and capture execution risks associated with cross-border transfer delays, while the last represents global blockchain activity.
    \item \textbf{CEX Benchmark Prices:} High-frequency (5-minute) BTC/USD prices from the Kraken exchange, serving as a highly liquid global benchmark price ($P^{\text{CEX}}_{\text{USD}}$).
    \item \textbf{Institutional Frictions:} Binary indicators of exchange rate regimes and capital controls extracted from the narratives in IMF's \textit{Annual Report on Exchange Arrangements and Exchange Restrictions (AREAER)}. 
    \item \textbf{Remittance Costs:} The percentage cost of sending and receiving \$500 across borders, obtained from the World Bank's \emph{Remittance Prices Worldwide} (RPW) database.\footnote{\url{http://remittanceprices.worldbank.org}, last accessed in Oct. 2024.} 
\end{itemize}

A comprehensive dataset dictionary and a table of variables with detailed extraction procedures and formal definitions are provided in \Cref{Appendix:data_dict}. While our primary analysis is conducted at the daily frequency, we also construct an auxiliary weekly dataset to accommodate lower-frequency heterogeneity analysis and to study the relationship between P2P premiums and OER depreciation. The aggregation procedure is described in \Cref{Appendix:data_aggregation}.

Specifically, we construct our primary dependent variables according to the structural decomposition established in \cref{eq:framework_log} (see \Cref{Appendix:variable_p2p}), including the total P2P premium, the baseline P2P spread, and the country-specific component---using aggregated local P2P prices, P2P USD prices, CEX benchmark prices, and OERs.

For the key explanatory variables capturing local macroeconomic conditions and global crypto market risks, we construct the following variables. First, to capture global crypto market conditions (see \Cref{Appendix:variable_crypto}), we compute BTC market returns using log-differenced prices from the Kraken benchmark, and proxy global BTC market volatility using daily realized volatility based on 1-hour sampled returns, calculated as the sum of squared hourly log returns over a specified rolling window. Second, we measure FX dynamics (see \Cref{Appendix:fx}). We define FX depreciation as the weekly log change in the OER, and measure FX volatility as the conditional standard deviation from a GARCH(1,1) model estimated using log exchange rate returns. For institutional constraints (see \Cref{Appendix:institutional}), we construct dummy variables indicating whether an economy operates under capital controls and whether its exchange rate regime is pegged. A combined dummy, $\text{Constrained}_{c,t}$, identifies economies subject to both constraints. Details on the construction and interpolation of remittance costs are provided in \Cref{sec:remittance_def}.

\subsection{Stylized Facts of the P2P Market}
Before proceeding to our formal empirical specifications, we document several stylized facts regarding the liquidity, distribution, and spatial heterogeneity of P2P BTC premiums. These patterns provide crucial intuition for our subsequent regression analyses.
\begin{figure}[b]
\centering

    \begin{subfigure}[b]{0.48\textwidth}
        \centering
        \includegraphics[width=\textwidth]{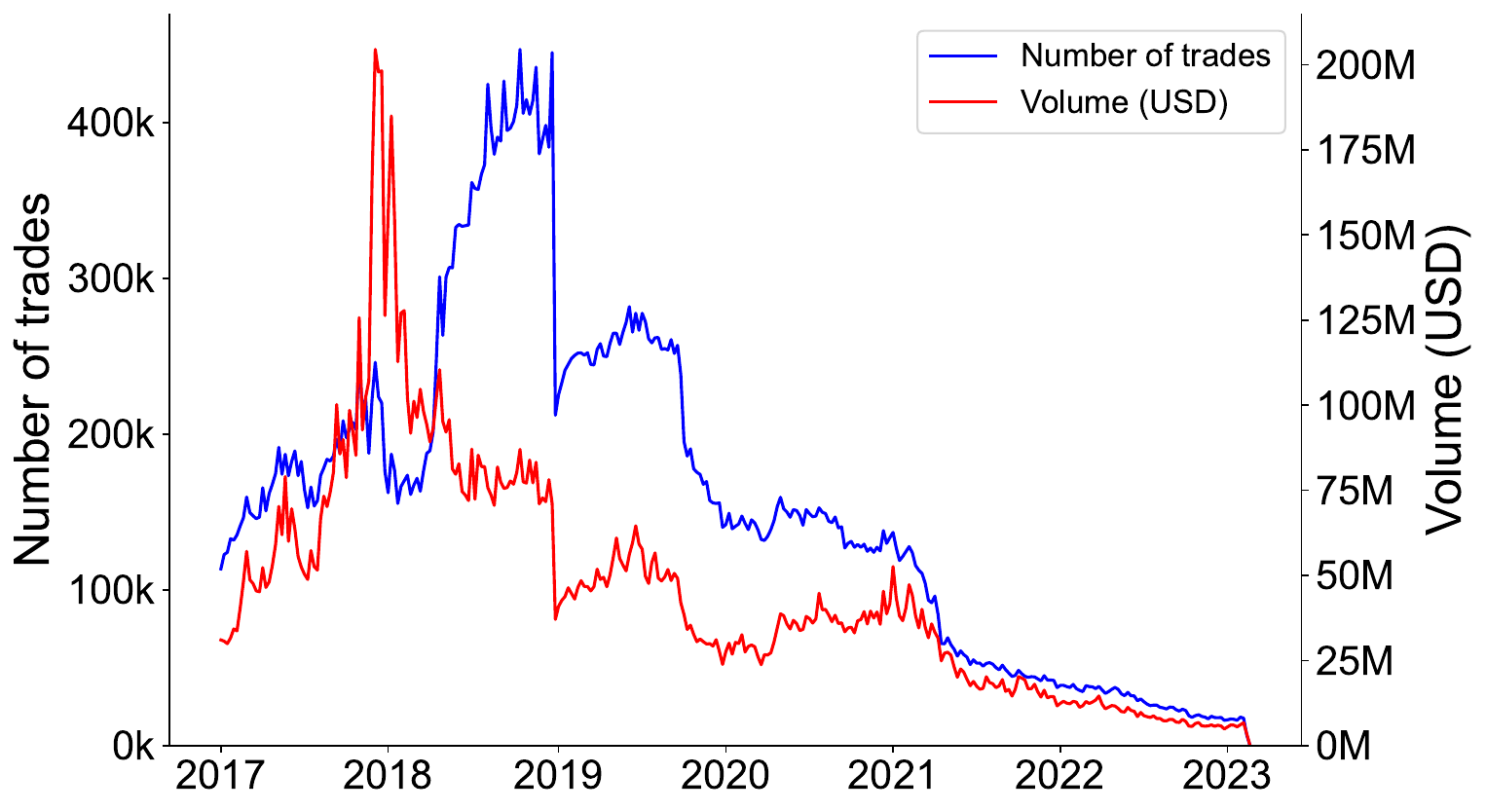}
        \caption{Weekly trading volume and number of transactions}
        \label{fig:trades}
    \end{subfigure}
    \hfill
    \begin{subfigure}[b]{0.48\textwidth}
        \centering
        \includegraphics[width=\textwidth]{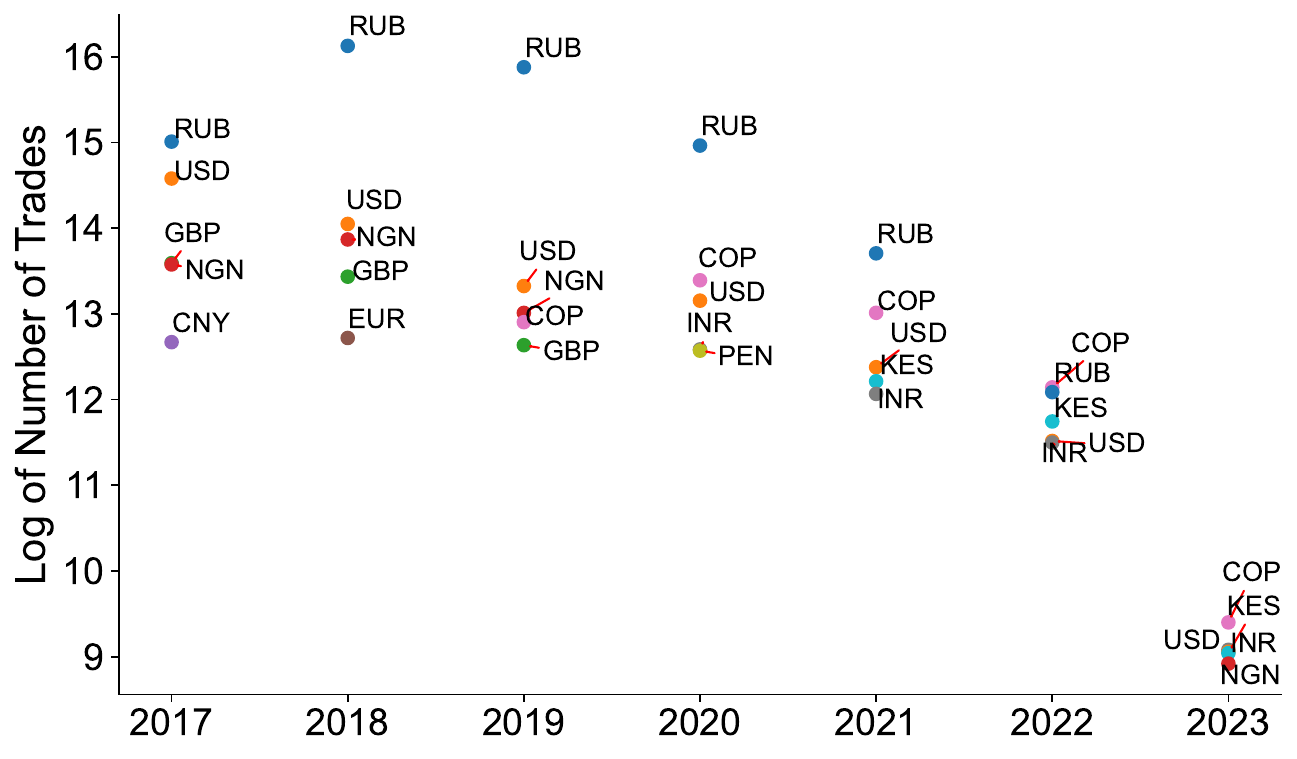}
        \caption{Popularity of currencies}
        \label{fig:currency}
    \end{subfigure}
    
    \caption{Trading Activity on LocalBitcoins, 2017--2023}
    \label{fig:tradesOverview}
    
    \begin{minipage}{0.96\textwidth}
    \footnotesize
    \textit{Notes:} Panel (a) reports weekly trading volume (in USD) and the number of transactions on LB. Panel (b) shows the top five fiat currencies by trade count in each year.
\end{minipage}

\end{figure}
\paragraph{Trading Activity and Market Characteristics.} 
We document a persistent divergence between trading volume and transaction counts, suggesting an increasing role of retail participation over time. Throughout our sample period, LB recorded substantial trading activity across many currencies, indicating that it was an economically meaningful alternative trading venue. As illustrated in \cref{fig:trades}, weekly trading volume peaked at approximately \$200 million in early 2018 and declined thereafter, while the total number of transactions continued to increase, reaching its peak toward the end of 2018. This divergence is consistent with a change in market composition: although speculative capital receded during the crypto bear market, transaction activity remained elevated, pointing to sustained usage by retail participants. In later periods, trading volume stabilized at a lower level, converging to a range of \$5.5 to \$7.5 million per week until the platform's closure.

The cross-sectional distribution of trading activity further highlights the platform's global reach. The top five most heavily traded fiat currencies---the Russian ruble (RUB), USD, British pound (GBP), Nigerian naira (NGN), and Colombian peso (COP)---span both developed and emerging markets (see \cref{fig:currency} for a yearly breakdown). The prominence of RUB is particularly notable, as it suggests that LB was used intensively not only in major global currencies, but also popular in markets where the domestic currency had limited international acceptance and where users faced non-trivial frictions in FX conversion and cross-border financial intermediation.

At the transaction level, the average trade size was approximately 0.046 BTC, indicating a predominantly retail-driven market. The presence of large trades, with sizes reaching up to 285 BTC (approximately 2.4 million USD), suggests that the platform maintained sufficient depth to accommodate sizable capital flows.

\begin{figure}[htbp]
    \centering
 
    \begin{subfigure}[b]{0.48\textwidth}
        \centering
        \includegraphics[width=\textwidth]{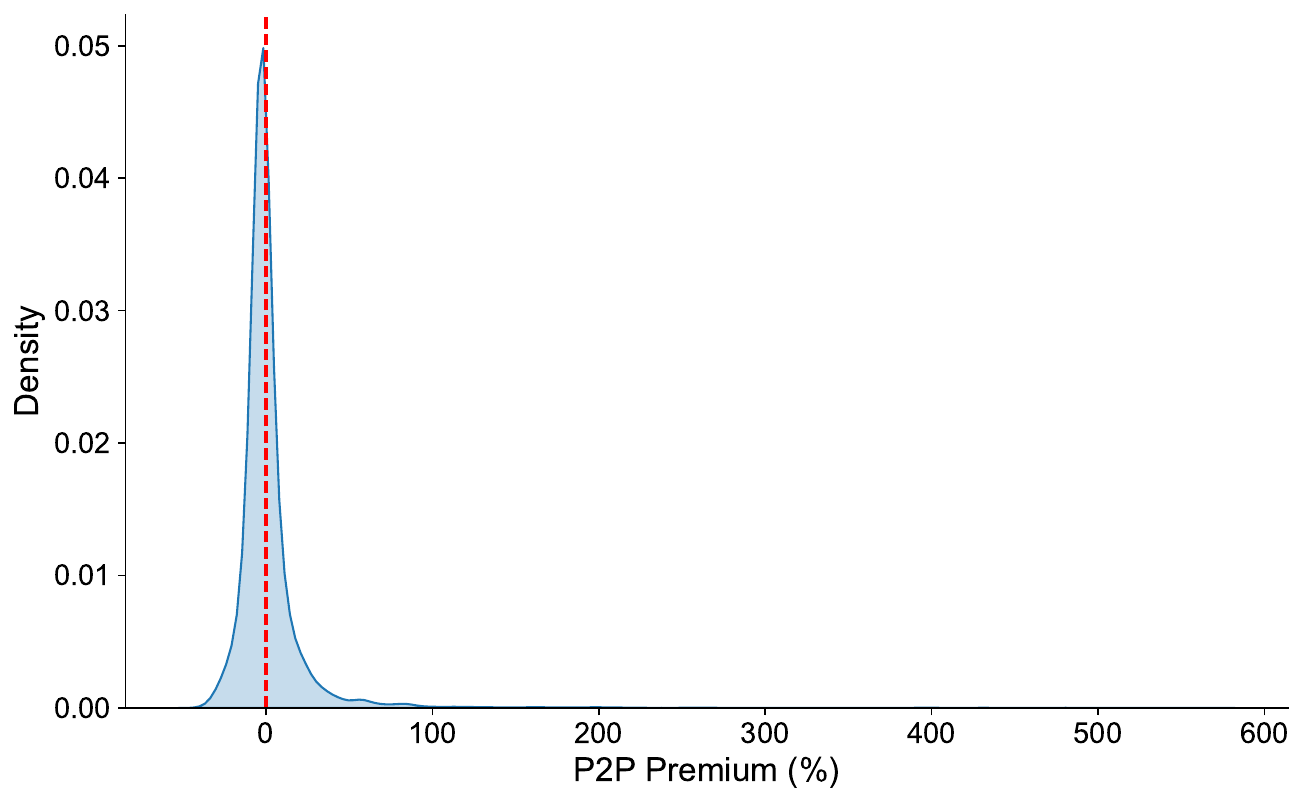}
        \caption{Pooled density of weekly premiums}
        \label{fig:Density}
    \end{subfigure}
    \hfill
    \begin{subfigure}[b]{0.48\textwidth}
        \centering
        \includegraphics[width=\textwidth]{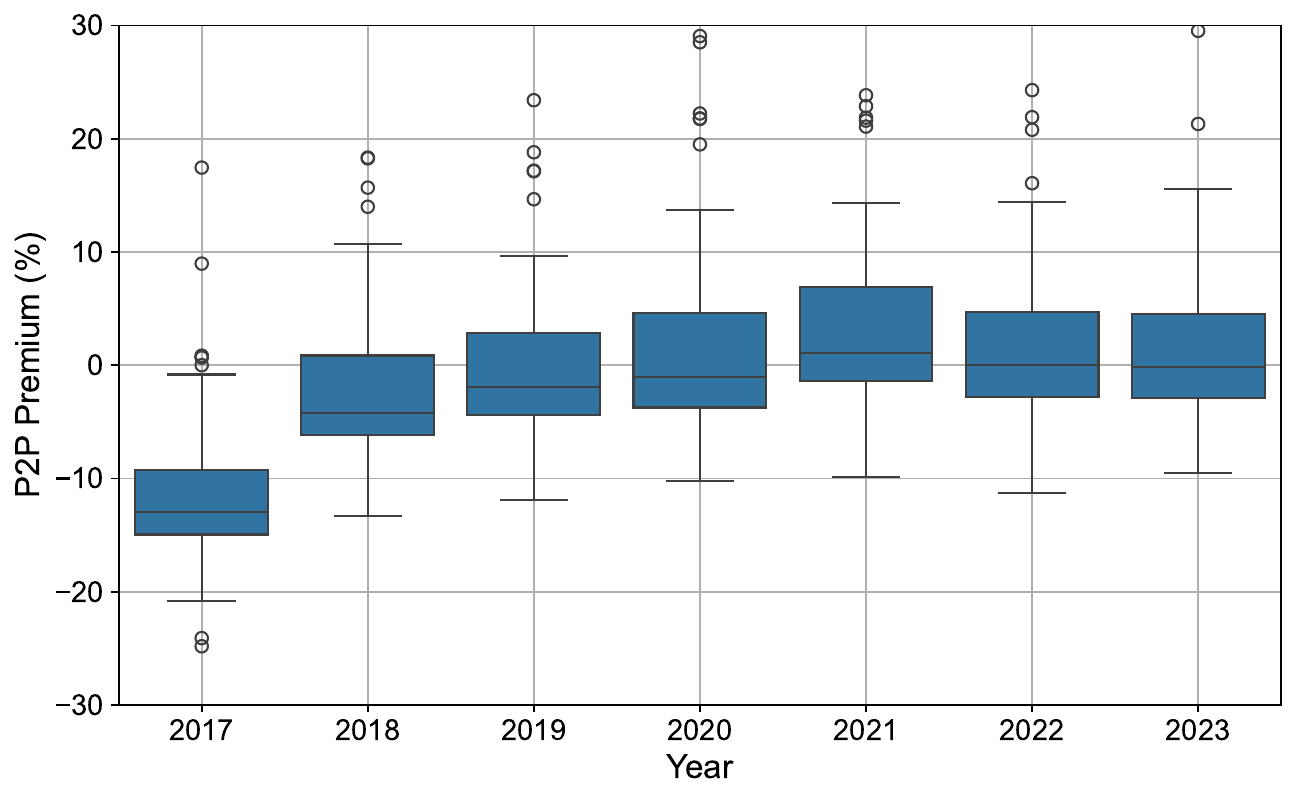}
        \caption{Annual distribution of premiums}
        \label{fig:premiums_yearly}
    \end{subfigure}
    
    \caption{Distribution and Dynamics of P2P BTC Premiums}
    \label{fig:premiumsOverview}

    \begin{minipage}{0.96\textwidth}
    \footnotesize
    \textit{Notes:} This figure presents the distribution of local P2P BTC premiums. Panel (a) illustrates the pooled density of weekly averaged premiums across all sample currencies and periods, highlighting the extreme right-skewness of price deviations. Panel (b) shows the annual distribution via box plots, where the boxes represent the interquartile range (25th to 75th percentiles) and the interior horizontal lines mark the medians. For visual clarity, the vertical axis is truncated at -30\% and +30\%, omitting a small number of extreme outliers. The upward shift in both the median and the interquartile range suggests a secular increase in global P2P premiums over the sample period.
    \end{minipage}
\end{figure}

\paragraph{Distribution and Temporal Dynamics of Premiums.} 
The pooled distribution of weekly BTC premiums is highly right-skewed (\cref{fig:Density}), with a long right tail. While the median premium is close to zero (-1.9\%), the distribution exhibits substantial asymmetry, with a maximum premium of 572.2\% compared to a minimum discount of -42.5\%. This pronounced right tail indicates the presence of large and infrequent deviations from parity. 

Temporally, we observe a transition from persistent discounts in the early period to predominantly positive premiums in later years (\cref{fig:premiums_yearly}). During the early sample period (2017--2018), premiums were predominantly negative despite high trading volumes. This early persistent discount suggests that users were willing to sell BTC below CEX benchmark prices in exchange for the platform's unique convenience yields---such as a wider array of localized payment options and, crucially, looser identity verification requirements prior to 2019.\footnote{Strict Know-Your-Customer (KYC) verifications, requiring ID document submission for volumes exceeding \$1,000 USD annually, were progressively enforced starting in 2019. See \url{https://localbitcoins.com/blog/id-verification-update/}, last accessed in Oct. 2024.} From 2021 onward, as regulatory scrutiny tightened, the global distribution shifted, with more than half of the currencies exhibiting positive annual premiums.

\begin{figure}
    \centering
    \includegraphics[height=0.58\textheight, angle=90]{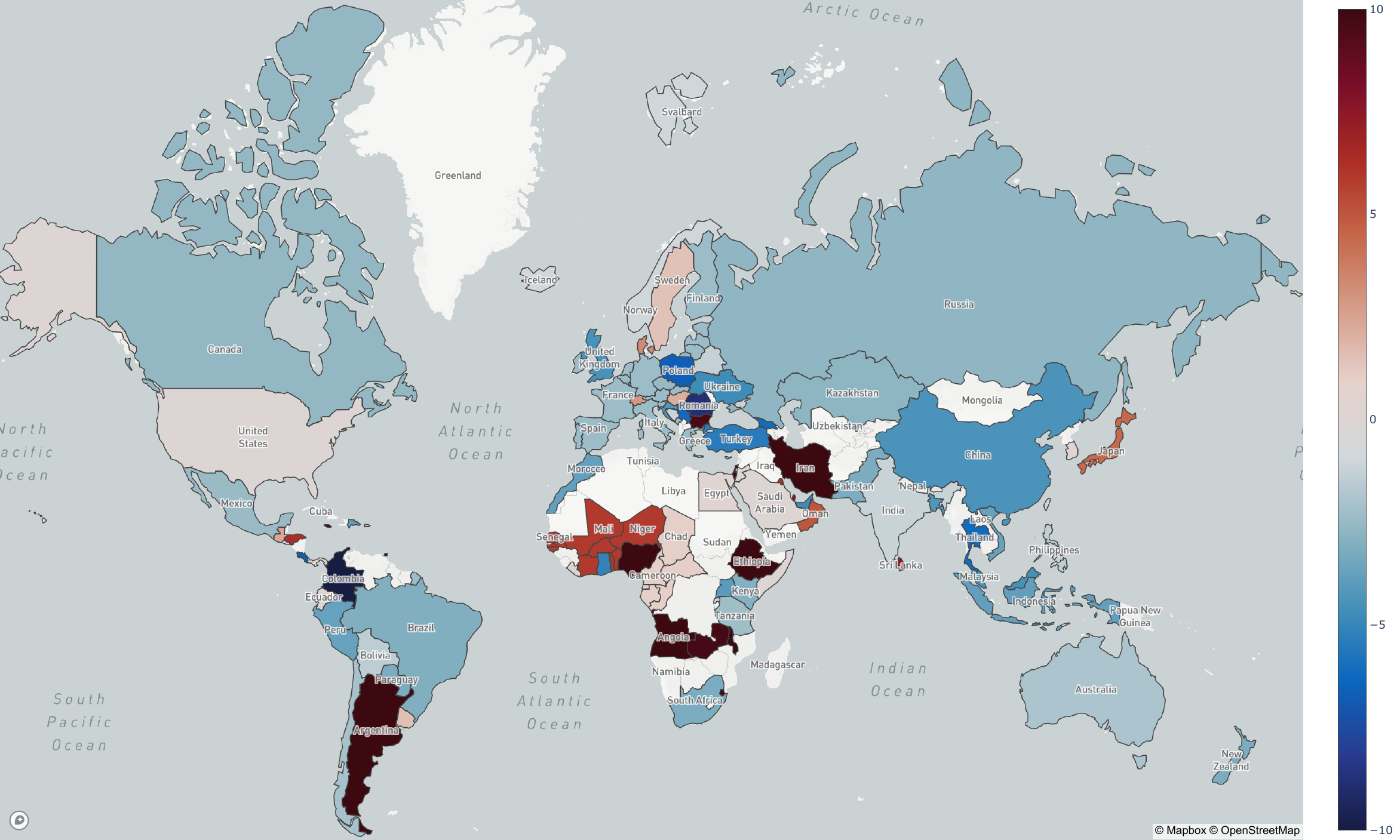}
    \caption{Median BTC Premiums by Country}
    \label{fig:Crosssection}
    
    \begin{minipage}{\textwidth}
    \footnotesize
    \textit{Notes:} The figure reports the median of weekly BTC premium for each country from January 2017 to February 2023. Premiums are first computed at the currency level and then aggregated to the country level, such that countries sharing the same currency exhibit identical values.
    \end{minipage}
\end{figure}

\paragraph{Spatial Heterogeneity and Institutional Constraints.} 
We document pronounced cross-country heterogeneity in BTC premiums, strongly associated with institutional and market frictions. \cref{fig:Crosssection} illustrates the geographic distribution of median BTC premiums across countries. 

A clear spatial pattern emerges. Developed economies in North America and parts of Europe tend to exhibit small discounts, whereas large positive premiums are concentrated in Sub-Saharan Africa, Latin America, and several Asian jurisdictions. Extreme values are primarily observed in economies with limited access to hard currency. For instance, Iran (IRR, 168.4\%), Ethiopia (ETB, 57.9\%), Angola (AOA, 26.9\%), and Argentina (ARS, 24.9\%) rank among the highest-premium countries. In contrast, persistent discounts are more common in relatively liquid P2P markets with better access to cross-border crypto trading, such as Colombia (COP, -10.1\%). Overall, positive premiums are more prevalent and more dispersed than discounts, consistent with asymmetric frictions across countries.

This spatial divide is further reflected in \cref{Tab:SummaryStatistics}, which compares weekly currency-level statistics across constrained and unconstrained groups, where constrained currencies are those subject to both capital controls and non-floating exchange rate regimes.

Constrained currencies exhibit substantially higher premiums (11\% vs. 1\%), along with greater exchange rate depreciations. Despite higher traditional remittance costs, they also maintain robust trading activity, as reflected in sustained weekly volumes. These patterns are consistent with P2P markets serving as an important channel for capital intermediation in environments with binding financial restrictions, underscoring the role of country-level institutional frictions.

\input{RegressionTables/SummaryStatistics}

\subsection{Representative Country Case Studies}
To concretize how institutional constraints shape P2P pricing dynamics, we examine four country case studies, illustrated in \cref{fig:CountryExamples}. Details about our full sample can be found in \Cref{Appendix:pairs}. 

\begin{figure}[!tbp]
    \centering
    
    \begin{subfigure}[b]{0.48\textwidth}
        \centering
        \includegraphics[width=\textwidth]{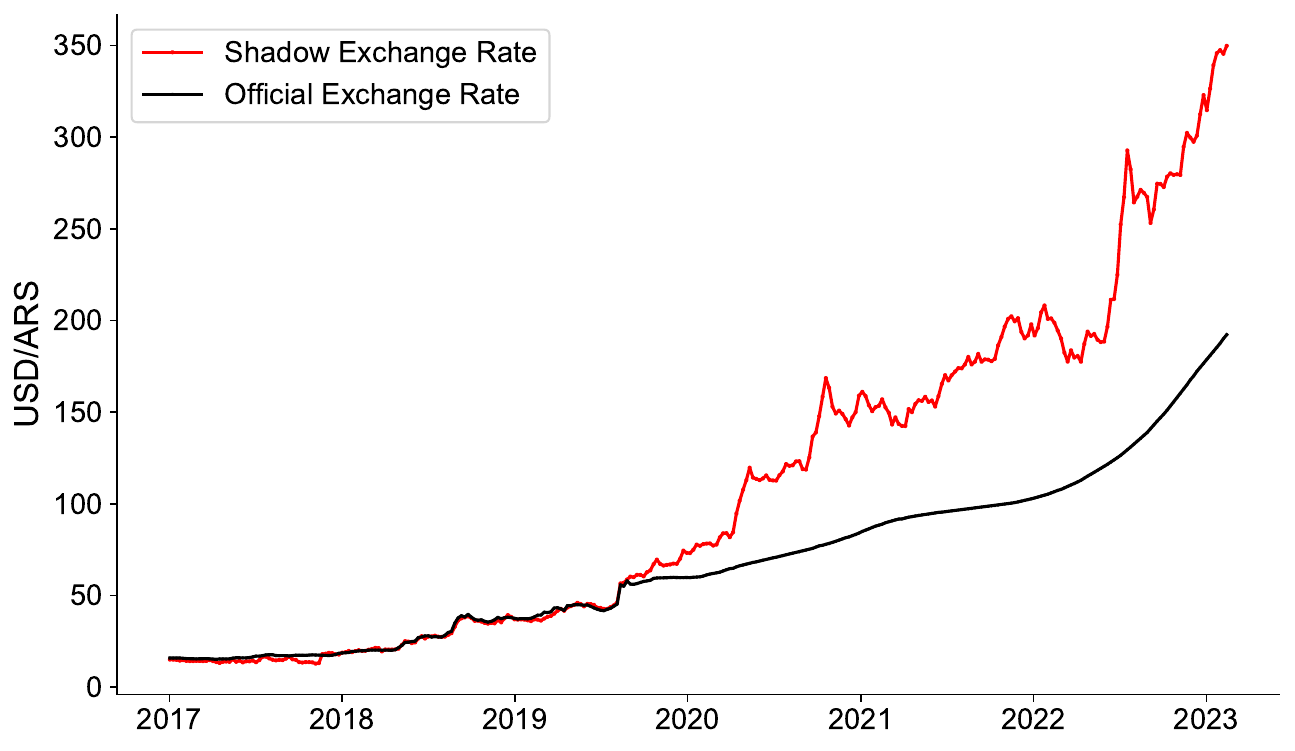}
        \caption{Argentina (ARS)}
        \label{fig:Argentina}
    \end{subfigure}\hfill
    \begin{subfigure}[b]{0.48\textwidth}
        \centering
        \includegraphics[width=\textwidth]{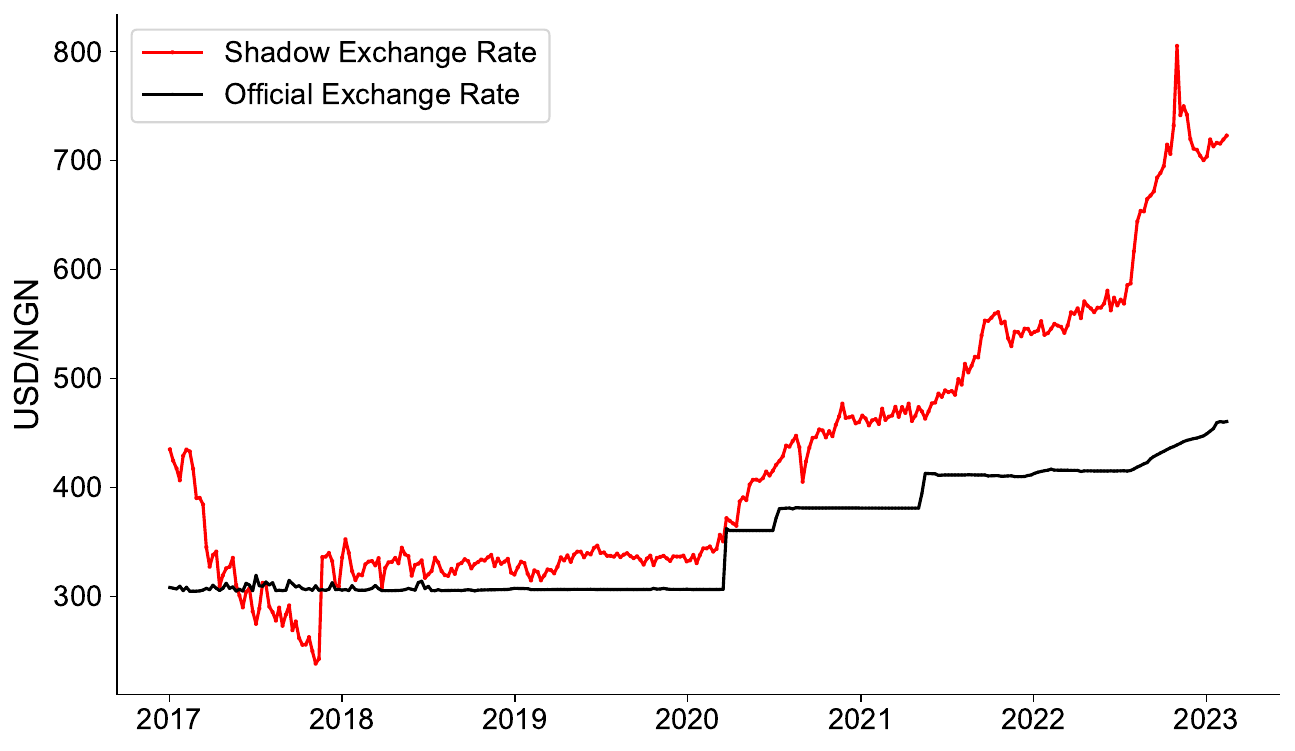}
        \caption{Nigeria (NGN)}
        \label{fig:Nigeria}
    \end{subfigure}
    
    \vspace{1em} 
    
    \begin{subfigure}[b]{0.48\textwidth}
        \centering
        \includegraphics[width=\textwidth]{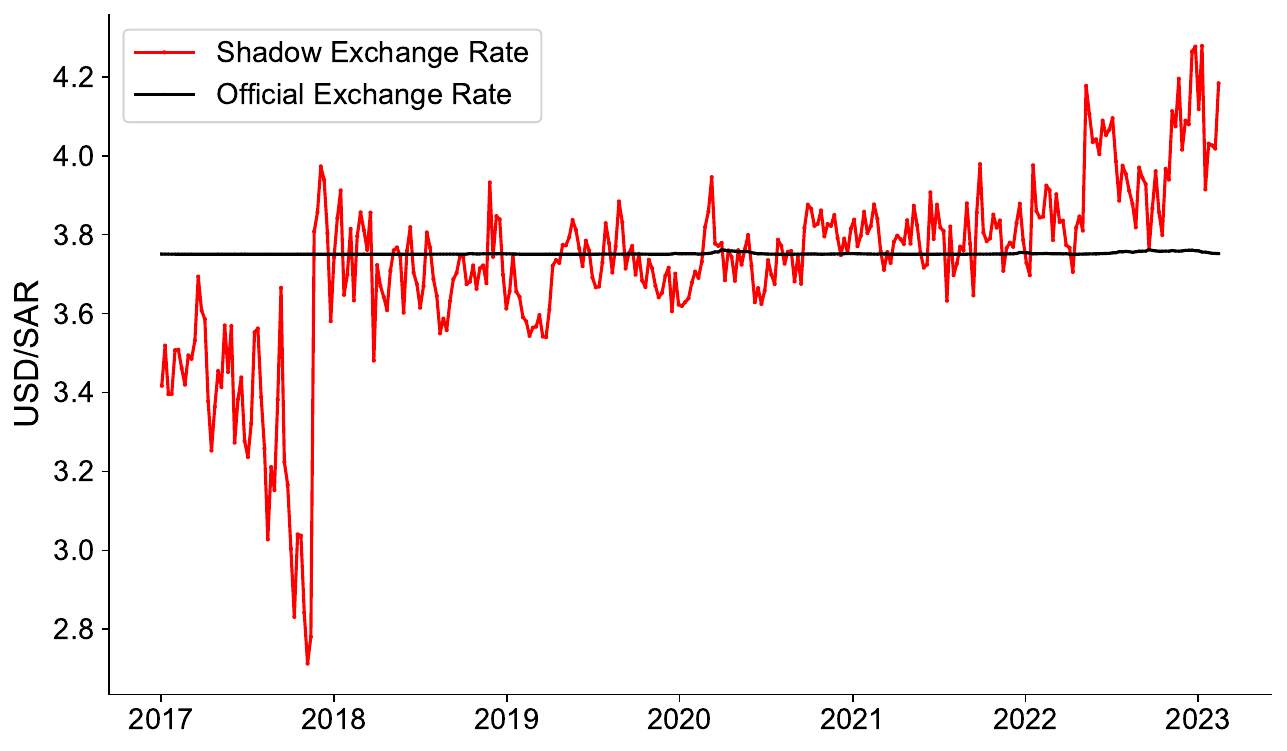}
        \caption{Saudi Arabia (SAR)}
        \label{fig:SaudiArabia}
    \end{subfigure}\hfill
    \begin{subfigure}[b]{0.48\textwidth}
        \centering
        \includegraphics[width=\textwidth]{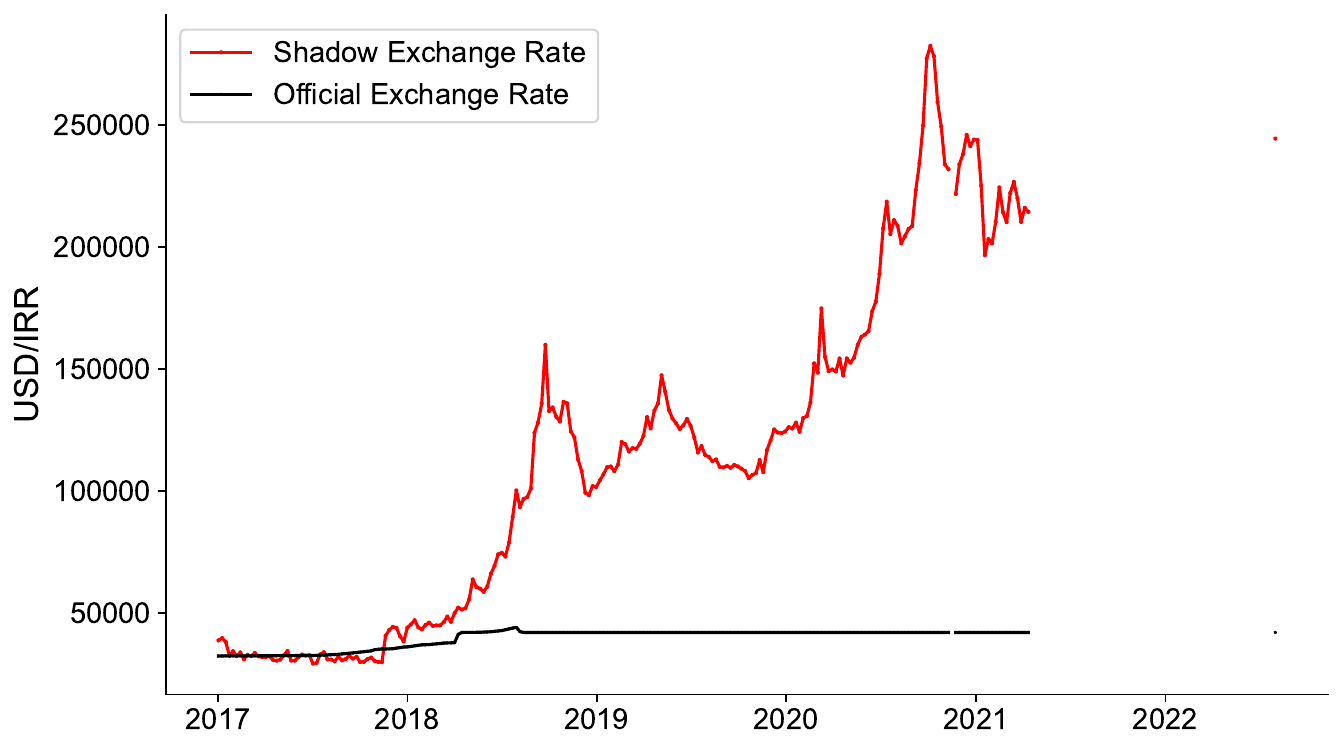}
        \caption{Iran (IRR)}
        \label{fig:Iran}
    \end{subfigure}
    
    \caption{Official vs. Shadow Exchange Rates: Selected Case Studies}
    \label{fig:CountryExamples}

    \begin{minipage}{\textwidth}
    \footnotesize
    \textit{Notes:} These figures illustrate the divergence between the OERs and the P2P SERs for four representative economies. Argentina and Nigeria demonstrate how capital controls and USD rationing lead to a severe decoupling of the SER, which preemptively signals subsequent official currency devaluations. Saudi Arabia and Iran offer a comparative static of officially pegged regimes: the unconstrained Saudi Riyal exhibits no premium, while the heavily sanctioned Iranian Rial experiences massive shadow pricing distortions. Note: Values are scaled according to their respective local currencies per USD.
    \end{minipage}

\end{figure}

\paragraph{Capital Controls and Devaluation Pressure: Argentina and Nigeria.} 
In \cref{fig:Argentina}, the Argentine SER closely tracks the OER prior to late 2019. In mid-2019, however, the government introduced strict capital controls to stabilize the currency and limit USD outflows, including restrictions on foreign currency purchases. Following these measures, the SER begins to diverge markedly from the OER. This divergence coincides with a sharp increase in P2P trading activity: average daily transactions rise from around 140 before 2019 to 382 afterward, eventually exceeding 800 per day by mid-2020. The widening gap between the SER and OER is consistent with growing devaluation pressure and increasing demand for alternative channels to access foreign currency.

Similarly, Nigeria (\cref{fig:Nigeria}) operates under a managed exchange rate regime. Around 2018, its P2P premium turns persistently positive. In July 2020, amid acute USD shortages, Nigerian banks impose strict limits on foreign currency access, including caps on international debit card spending below \$100 per month.\footnote{See \url{https://www.reuters.com/article/idUSKCN24K0E9/}, last accessed in Mar. 2026.} Following this tightening of FX access, deviations between the P2P SER and the official rate increase substantially. Notably, the persistent divergence precedes two major devaluations of the official NGN rate, suggesting that P2P prices incorporate forward-looking information about exchange rate pressures.

\paragraph{Pegged Exchange Rate Regimes: Saudi Arabia vs. Iran.}
\cref{fig:SaudiArabia} and \cref{fig:Iran} provide a contrast between two economies operating under OERs anchored to the USD. Saudi Arabia maintains a highly credible peg with minimal restrictions on cross-border capital flows. Consistent with a low-friction environment, its SER closely tracks the OER, with only limited deviations. In contrast, Iran also maintains an official peg but is subject to severe international sanctions and tight capital controls. In this setting, the SER exhibits a large and persistent divergence from the OER, reflecting the underlying macroeconomic pressures that are not captured by the official rate.\footnote{The data for Iran concludes in mid-2021 when LB operations were officially banned in the country.}

Taken together, these examples highlight the role of institutional frictions in shaping cross-market price discrepancies. In environments with binding constraints, the P2P premium behaves as a persistent wedge relative to the OER. Moreover, the observed dynamics are consistent with the P2P SER incorporating forward-looking information about exchange rate pressures. We formally test this predictive relationship in \cref{sec:var}.

%% file: RegressionTables/transactions.tex
\begin{table}[!tbp]
\centering
\caption{Examples of Historical Trades}
\label{Tab:tab_transaction}

\begin{threeparttable}
\renewcommand{\arraystretch}{1.1}

\begin{tabular}{lcccc}
\toprule
Timestamp & Currency & Transaction ID & Price & Amount \\
\midrule
1649809649 & EUR & 55600084 & 40579.86 & 0.00418927 \\
1649743308 & CHF & 55593671 & 34998.07 & 0.00145208 \\
\bottomrule
\end{tabular}

\begin{tablenotes}[flushleft]
\footnotesize
\item \textit{Notes:} This table reports illustrative transactions from LB. Each observation records the transaction timestamp (in Unix time), the currency in which BTC was sold, the transaction identifier, the transaction price per BTC, and the amount of BTC sold.
\end{tablenotes}

\end{threeparttable}
\end{table}

%% file: RegressionTables/SummaryStatistics.tex
\begin{table}[!tbp]
\centering

\begin{threeparttable}
\caption{Summary Statistics by Institutional Constraint}
\label{Tab:SummaryStatistics}

\small
\setlength{\tabcolsep}{5pt}
\renewcommand{\arraystretch}{1.1}

\begin{tabular}{lrrrrrrrr}
\toprule
Variable & \multicolumn{1}{c}{N} & \multicolumn{1}{c}{Mean} & \multicolumn{1}{c}{Std. Dev.} & \multicolumn{1}{c}{Min} & \multicolumn{1}{c}{Pctl. 25} & \multicolumn{1}{c}{Pctl. 50} & \multicolumn{1}{c}{Pctl. 75} & \multicolumn{1}{c}{Max} \\
\midrule
\multicolumn{9}{l}{\textit{Constrained = 0}} \\
\addlinespace
Premium (\%) & 17796 & 0.99 & 29 & -41 & -6.6 & -2.3 & 2.7 & 571 \\ 
Depreciation (\%) & 17798 & 0.028 & 1.2 & -24 & -0.38 & 0 & 0.4 & 24 \\ 
P2P Volume (BTC) & 17798 & 72 & 309 & 0.00014 & 0.44 & 3.8 & 34 & 5396 \\ 
Remittance Cost \% & 10576 & 4.5 & 2.3 & 0 & 2.9 & 4.2 & 5.7 & 23 \\ 
\addlinespace
\multicolumn{9}{l}{\textit{Constrained = 1}} \\
\addlinespace

Premium (\%) & 3093 & 11 & 32 & -41 & -4.9 & -0.6 & 13 & 322 \\ 
Depreciation (\%) & 3093 & 0.096 & 1 & -5.8 & -0.086 & 0.0027 & 0.22 & 26 \\ 
P2P Volume (BTC) & 3093 & 48 & 224 & 0.0001 & 0.23 & 2.4 & 11 & 4266 \\ 
Remittance Cost \% & 1939 & 5.3 & 3.1 & 0 & 3.1 & 4.5 & 6.6 & 20\\ 
\bottomrule
\end{tabular}

\begin{tablenotes}[flushleft]
\footnotesize
\item \textit{Notes:} The table reports weekly summary statistics by institutional constraint status, defined as in \cref{eq:Constrained} with threshold $\sigma = 0.7$. Variables include BTC premium, exchange rate depreciation, trading volume (BTC), and remittance costs (\%).
\end{tablenotes}

\end{threeparttable}
\end{table}

%% file: Sections/Result_2026.tex
\section{Empirical Results and Mechanism Analysis}\label{sec:results}
According to \cref{eq:framework}, we structurally decompose the total P2P premium into a global USD P2P baseline friction and a local currency-specific spread. We then regress the total premium and its two underlying components separately on the same set of explanatory variables to identify specific transmission channels.  

\subsection{Baseline Results and Heterogeneity Analysis}\label{sec:base}
\subsubsection{Base Model Results}\label{sec:base_model}
\cref{Tab:p2p_base} reports the estimation results of the baseline dynamic panel model in \cref{eq:prem_base}. We regress the total P2P premium (Column 1) and its two structural components---the baseline USD friction (Column 2) and the currency-specific spread (Column 3)---on a common set of market and institutional variables. To preserve the exact linear additivity of the OLS estimator (e.g. $\hat{\beta}_{\text{total}} = \hat{\beta}_{\text{baseline}} + \hat{\beta}_{\text{specific}}$), all three specifications control for the identical lagged total premium.

\input{RegressionTables/base}

\paragraph{Persistence}
The lagged premium exhibits a strong positive effect across specifications, indicating substantial persistence in P2P price deviations. This persistence is more pronounced in the currency-specific component, suggesting that local market frictions tend to evolve slowly over time. By contrast, deviations in the USD benchmark market are comparatively short-lived, which may reflect higher liquidity and more effective cross-market arbitrage.

\paragraph{Trading Volume and Liquidity}
The results indicate that trading volume reduces frictions in both the baseline USD market and local currency markets, but the magnitude of the effect differs across the two. Given the high liquidity of the USD benchmark market, additional volume leads to only modest reductions in baseline frictions. In contrast, local markets are typically less liquid and more fragmented. Increases in local trading volume are associated with greater participation and competition, which compresses the currency-specific spread. As a result, improvements in local liquidity are associated with a decline in the overall relative P2P premium.

\paragraph{Blockchain Network Fundamentals}
On-chain frictions, measured by on-chain costs (in USD) and median confirmation times (in minutes), exhibit heterogeneous effects across the baseline USD component and the local currency-specific component. 

Empirically, higher on-chain frictions, including median formation time and on-chain costs, are associated with an expansion of the baseline USD spread, while the currency-specific deviation declines, resulting in a contraction of the overall premium. Within the framework in \cref{eq:framework}, the CEX USD price, $P^{\text{CEX}}_{\text{USD}}$, serves as a common reference that is plausibly less affected by these specific on-chain frictions, suggesting differential price adjustment across the baseline USD and local-currency segments of the P2P market. In more liquid USD-denominated P2P markets, intermediaries may be better positioned to pass increased transaction costs onto buyers. In contrast, local markets appear more price-sensitive, and adjustments in local prices are more limited. Consistent with this interpretation, higher on-chain costs are also associated with reduced trading activity in local markets (see \cref{sec:amount} for further evidence), suggesting that participants may respond to higher costs by delaying or reducing transactions.

Regarding overall blockchain network activity, it has a negative effect on baseline USD frictions ($-0.086^{***}$) and a smaller negative effect on local deviations ($-0.021^{***}$), resulting in a positive net effect on the relative P2P premium. Periods of elevated network activity are associated with greater global liquidity, facilitating flows between P2P markets and other trading channels, and compressing spreads, particularly in the USD baseline market.



\paragraph{BTC Market Dynamics}
A broad appreciation in BTC's USD price on CEXs, as captured by BTC returns, is associated with an increase in both the USD baseline and the currency-specific spread, consistent with rising BTC prices driving demand in P2P markets across both segments. The effect is more pronounced in the local currency-specific component, suggesting that the same demand pressure generates a larger price impact in local markets, where liquidity is lower and the pool of counterparties is more limited.

Short-term BTC volatility, by contrast, is associated with a decline in the overall premium ($-0.088^{***}$). Higher volatility widens baseline USD spreads ($0.172^{***}$), while the local component responds more weakly. Because the baseline rises by more than the local component, the relative premium narrows. This asymmetry is consistent with local P2P markets being more isolated from global order flow. Thus, high-frequency risk signals transmit more rapidly in P2P USD market that is more closely integrated with global crypto markets.

\paragraph{Macroeconomic and FX Frictions}
OER volatility has a strong negative association with the total P2P premium ($-0.726^{**}$). This effect operates almost entirely through the country-specific component ($-0.745^{*}$), with little impact on the USD baseline. To further examine the role of FX frictions, we extend the baseline specification by introducing institutional dummies and their interaction terms; see \cref{sec:dummy} for details.

\paragraph{Synthesis of Mechanism Asymmetries}
Overall, the results point to a structural asymmetry between the baseline USD component and the local currency-specific component. Variables related to network conditions and short-term BTC risk (volatility) are more strongly reflected in the USD baseline component, whereas P2P trading volume, BTC returns, and FX volatility play a more prominent role in shaping currency-specific spreads.

These patterns highlight the importance of distinguishing between global and local drivers of P2P premiums. In particular, factors related to blockchain network conditions and short-term BTC volatility tend to be more strongly transmitted through the P2P USD market, while demand-driven BTC price movements and local macroeconomic conditions, such as FX volatility, are more closely linked to currency-specific spreads. A more detailed analysis of institutional frictions is provided in \cref{sec:dummy,sec:remittance}.

\subsubsection{Macro Heterogeneity: The Role of Institutional Constraints}\label{sec:dummy}
\input{RegressionTables/2dummy0.7}
To isolate heterogeneity in the effect of FX volatility on P2P premiums, we extend the baseline specification to \cref{eq:hetero_inst} by interacting FX volatility with a capital control dummy, $\mathrm{CC}_{c,t}$, and an exchange rate regime dummy, $\mathrm{ERR}_{c,t}$. Here, $\mathrm{CC}_{c,t}=1$ indicates that the economy is classified as severe capital controls, and $\mathrm{ERR}_{c,t}=1$ indicates a non-floating exchange rate regime; the construction of both indicators is described in \cref{eq:cc,eq:ERR}. It should be noted that $\mathrm{ERR}_{c,t}=1$ does not imply an absence of FX volatility; managed regimes frequently can still exhibit FX fluctuations, as interventions are typically aimed at smoothing excessive volatility rather than maintaining a fixed parity (\Cref{Appendix:institutional}). The results are reported in \cref{Tab:2dummy0.7}. We focus on the total premium and the currency-specific spread because the institutional mechanism of interest is expected to operate primarily through local market frictions rather than through the global USD baseline component.

The baseline coefficient on official FX volatility in the full model (Column 3) is significantly negative ($-0.951^{**}$). In this interaction specification, the main effect on FX volatility captures the association for economies with $\mathrm{CC}_{c,t}=0$ and $\mathrm{ERR}_{c,t}=0$, i.e., economies without capital controls and with floating exchange rate regimes. This suggests that, in less constrained environments, higher official FX volatility is associated with a lower P2P premium. One possible interpretation is that when exchange rates adjust more flexibly and capital can move through alternative formal channels, pressure is less likely to concentrate in the P2P market alone.

The interaction terms $\text{CC} \times \text{FX volatility}$ and $\text{ERR} \times \text{FX volatility}$ are not statistically significant in either the separate interaction models (Columns 1 and 2) or the full model (Column 3). This indicates that neither capital controls nor exchange rate management alone significantly alters the baseline relationship between FX volatility and the P2P premium. Economically, when only one constraint is binding, the other channel remains open for adjustment: a freely floating exchange rate can absorb currency pressure without driving demand into P2P markets, and the absence of capital controls leaves formal cross-border channels available, reducing incentives to route transactions through P2P markets. A single binding constraint is therefore insufficient to amplify the pass-through of FX volatility into P2P premiums.

In contrast, the triple interaction term, $\text{CC} \times \text{ERR} \times \text{FX volatility}$, is positive and statistically significant ($6.390^{**}$). This result implies that in countries characterized by both capital controls and managed exchange rates, the effect of FX volatility on the P2P premium is substantially amplified. Economically, this amplification reflects the simultaneous closure of the two primary adjustment channels. Under a managed regime, exchange rates adjust only partially to absorb changes in FX conditions, while capital controls restrict access to foreign currency through formal channels. Consequently, FX volatility leads to a stronger increase in demand in the P2P market, where prices adjust more freely, resulting in a larger premium.

Taken together, these findings suggest that large P2P price deviations arise primarily when exchange rates do not adjust freely and cross-border capital movements are restricted. Under these conditions, fluctuations in the FX markets are more likely to be reflected in P2P Bitcoin prices. Having established that access to formal cross-border financial channels plays a key role in shaping P2P pricing dynamics, we now proceed to test this substitution mechanism directly in \cref{sec:remittance}.

\subsubsection{Substitution Channel: Evidence from Remittance Costs}\label{sec:remittance}
\input{RegressionTables/remittance_weekly}
To test the substitution channel between traditional financial systems and P2P crypto markets, we introduce cross-border remittance costs ($\text{RC}$) as a proxy for frictions in traditional financial channels, and examine how these costs affect P2P pricing under varying institutional constraints. To better adapt to the frequency of the remittance data, we aggregate our sample to a weekly level (See \Cref{Appendix:data_aggregation}). Additionally, we introduce a new variable, $\text{FX depreciation}$ to capture directional exchange rate movements within the baseline specification. Following \cref{sec:dummy}, we construct an indicator variable, $\text{Constrained}$, to capture the joint presence of capital controls and exchange rate management. High-frequency variables about blockchain network conditions are not included in this specification. The corresponding results are reported in \cref{Tab:remittance}, using the specification in \cref{eq:hetero_trans}.

The results reveal a clear asymmetry. In unconstrained environments ($\text{Constrained} = 0$), the baseline coefficient on $\text{RC}$ is statistically insignificant for both the total P2P premium ($-0.101$) and the country-specific deviation ($-0.121$). This suggests that in settings with relatively free capital mobility and diversified financial channels, marginal fluctuations in traditional remittance costs do not significantly impact P2P pricing. In such environments, the P2P cryptocurrency market functions as an alternative channel without commanding an additional friction-induced premium.

In contrast, a pronounced asymmetry emerges in constrained environments. The interaction term, $\text{Constrained} \times \text{RC}$, is significantly positive for both the total premium ($0.808^{**}$) and the currency-specific spread ($0.917^{**}$). This pattern is consistent with a \textit{substitution channel effect}, whereby increases in traditional remittance costs are associated with higher P2P premiums. Given our log-log specification with $\log(1+\text{RC})$, the estimated coefficient can be interpreted as the elasticity of the P2P premium with respect to the gross remittance cost. For instance, an increase in remittance costs from 5\% to 10\% (i.e., from 1.05 to 1.10 in gross terms) implies an increase of approximately 4.8\% in $1+\text{RC}$, which corresponds to an estimated 3.9\% increase in the P2P premium when $\beta = 0.808^{**}.$ One interpretation is that when access to conventional cross-border channels becomes more costly or restricted, a spillover effect is generated toward borderless BTC-based P2P transactions, allowing prices in these markets to adjust upward. The results are robust to alternative measures of remittance costs based on sending and receiving fees (see \Cref{Appendix:robust}).

While official FX volatility is not statistically significant across these specifications (Columns 1--3), OER depreciation is associated with a decline in measured premiums ($-0.252^{***}$). This pattern is consistent with forward-looking dynamics in P2P pricing, as depreciation of the OERs implies convergence toward SERs, reducing the observed premium.

Overall, these findings reinforce the hypothesis that the P2P cryptocurrency market functions as a substitute channel when traditional financial intermediation is constrained, with frictions in formal remittance channels positively spilling over into P2P pricing. While the preceding analysis establishes how institutional frictions shape P2P pricing, a complete understanding of market clearing requires examining the corresponding quantity dynamics. We now turn to examine how institutional and market constraints affect trading volume in P2P markets.

\subsubsection{Frictions and Trading Activity: Evidence on Volume Dynamics}\label{sec:amount}
\input{RegressionTables/amount}

Turning to the quantity dynamics shown in \cref{Tab:amount}, the baseline results indicate that P2P market liquidity is fundamentally driven by the endogenous characteristics of the underlying crypto-asset. Most notably, BTC hourly volatility enters with a positive and highly significant coefficient across all specifications. This strong positive association is consistent with the well-established volume--volatility relationship in the financial microstructure literature \citep{tauchen1983price}, suggesting that speculative trading activity intensifies during periods of heightened crypto-market uncertainty. Furthermore, the positive and significant coefficients on lagged P2P trading volume ($0.465^{***}$) and BTC returns ($0.495^{***}$) point to substantial persistence and momentum in local market participation. When controlling for blockchain network fundamentals (Column 3), the positive association with overall on-chain transactions ($1.347^{***}$) confirms that local P2P trading remains structurally anchored to broader network activity.

Within the network variables, confirmation delays and transaction costs have different empirical effects. Median confirmation time is statistically insignificant, whereas on-chain transaction costs have a significantly negative effect on trading volume. This pattern suggests that P2P participants are less sensitive to settlement delays than to explicit transaction costs. It also complements our earlier premium results: higher on-chain costs are associated with lower trading activity and a narrower currency-specific spread, consistent with reduced trading demand rather than stronger price pass-through.

Crucially, the volume dynamics differ from the premium dynamics in response to FX volatility. Official FX volatility, which is associated with higher premiums in constrained environments, does not have a statistically significant effect on trading volume. Even in settings characterized by institutional constraints (Column 2), the interaction term between the constrained indicator and FX volatility remains insignificant. This discrepancy points to limited supply responsiveness in local P2P markets. During periods of macroeconomic turbulence, increases in demand for crypto-assets are not matched by corresponding expansions in local inventory or liquidity provision. As a result, FX-related macroeconomic pressures in these environments do not translate into meaningful quantity adjustments, but are instead reflected primarily in price movements.

Overall, the analysis highlights an asymmetry in how local P2P markets adjust to different types of variation. Trading volume responds strongly to crypto-market returns and volatility, but shows limited adjustment to FX volatility, particularly under institutional constraints. Combined with the effect of on-chain transaction costs, this pattern points to persistent liquidity frictions in local markets. In constrained environments, these pressures appear to be reflected more in prices (premiums) than in quantities (trading volume).

%% file: RegressionTables/base.tex
\begin{table}[!tbp]
\centering
\begin{threeparttable}
\caption{Baseline Decomposition of P2P Premiums}
\label{Tab:p2p_base}

\renewcommand*{\arraystretch}{1.2}
\setlength{\tabcolsep}{5pt}

\begin{tabular}{lccc}
\toprule
 & P2P premium & Baseline P2P friction & Currency-specific spread \\
\cmidrule(lr){2-2}\cmidrule(lr){3-3}\cmidrule(lr){4-4}
 & (1) & (2) & (3) \\
 
\midrule

Prem$_{t-1}$             & $.571^{***}$  & $-.022^{***}$ & $.550^{***}$  \\
                         & $(.102)$      & $(.005)$      & $(.107)$      \\
\addlinespace[0.25em]

P2P trading volume       & $-.009^{***}$ & $-.002^{***}$ & $-.011^{***}$ \\
                         & $(.001)$      & $(.000)$      & $(.002)$      \\
\addlinespace[0.25em]

Median confirmation time & $-.010^{***}$ & $.006^{***}$  & $-.004^{***}$ \\
                         & $(.001)$      & $(.000)$      & $(.001)$      \\
\addlinespace[0.25em]

On-chain cost (USD)      & $-.038^{***}$ & $.028^{***}$  & $-.009^{**}$  \\
                         & $(.004)$      & $(.001)$      & $(.004)$      \\
\addlinespace[0.25em]

On-chain transactions    & $.065^{***}$  & $-.086^{***}$ & $-.021^{***}$ \\
                         & $(.005)$      & $(.001)$      & $(.006)$      \\
\addlinespace[0.25em]

BTC return               & $.037^{***}$  & $.432^{***}$  & $.470^{***}$  \\
                         & $(.011)$      & $(.001)$      & $(.011)$      \\
\addlinespace[0.25em]

BTC hourly volatility    & $-.088^{***}$ & $.172^{***}$  & $.088^{***}$  \\
                         & $(.013)$      & $(.004)$      & $(.014)$      \\
\addlinespace[0.25em]

FX volatility            & $-.726^{**}$  & $-.016$       & $-.745^{*}$   \\
                         & $(.360)$      & $(.048)$      & $(.399)$      \\

\midrule

Num. obs.               & $134922$ & $134936$ & $134936$ \\
Num. groups: currency   & $80$        & $80$        & $80$        \\
Num. groups: week       & $313$       & $313$       & $313$       \\

R$^2$ (within)          & $.356$      & $.217$      & $.378$      \\

\bottomrule

\end{tabular}

\begin{tablenotes}[flushleft]
\footnotesize
\item \textit{Notes:} This table reports panel regression estimates corresponding to \cref{eq:prem_base}. The dependent variables are the log total P2P premium (Column 1), the log baseline USD P2P friction (Column 2), and the log country-specific spread (Column 3). All specifications include the first lag of the total premium. All regressions include country and time fixed effects. Within R$^2$ is computed after removing fixed effects. Standard errors are clustered at the currency level and reported in parentheses.
\item $^{***}p<0.01$, $^{**}p<0.05$, $^{*}p<0.1$.
\end{tablenotes}

\end{threeparttable}
\end{table}

%% file: RegressionTables/2dummy0.7.tex
\begin{table}[!tbp]
\centering
\begin{threeparttable}
\caption{Determinants of P2P Premiums: Institutional Frictions and FX Volatility}
\label{Tab:2dummy0.7}

\renewcommand*{\arraystretch}{1.3}
\setlength{\tabcolsep}{5pt}

\begin{tabular}{lcccc}
\toprule
 & \multicolumn{3}{c}{P2P premium} & Currency-specific spread \\
\cmidrule(lr){2-4} \cmidrule(lr){5-5}
 & (1) & (2) & (3) & (4) \\
\midrule

\addlinespace
\multicolumn{5}{l}{\textit{Panel A: Baseline Controls}} \\
\addlinespace

Prem$_{t-1}$            & $.568^{***}$ & $.566^{***}$ & $.548^{***}$ & $.524^{***}$ \\
                        & $(.106)$     & $(.107)$     & $(.108)$     & $(.115)$     \\
\addlinespace[0.25em]

Median confirmation time & $-.009^{***}$ & $-.009^{***}$ & $-.009^{***}$ & $-.003^{**}$ \\
                        & $(.001)$      & $(.002)$      & $(.001)$      & $(.001)$     \\
\addlinespace[0.25em]

On-chain cost (USD)     & $-.038^{***}$ & $-.038^{***}$ & $-.039^{***}$ & $-.011^{***}$ \\
                        & $(.004)$      & $(.004)$      & $(.004)$      & $(.004)$     \\
\addlinespace[0.25em]

On-chain transactions   & $.063^{***}$  & $.064^{***}$  & $.066^{***}$  & $-.018^{***}$ \\
                        & $(.005)$      & $(.005)$      & $(.005)$      & $(.006)$     \\
\addlinespace[0.25em]

P2P trading volume      & $-.009^{***}$ & $-.009^{***}$ & $-.010^{***}$ & $-.012^{***}$ \\
                        & $(.001)$      & $(.001)$      & $(.001)$      & $(.001)$     \\
\addlinespace[0.25em]

BTC return              & $.041^{***}$  & $.042^{***}$  & $.042^{***}$  & $.468^{***}$ \\
                        & $(.011)$      & $(.012)$      & $(.012)$      & $(.012)$     \\

\addlinespace[0.6em]
\multicolumn{5}{l}{\textit{Panel B: Institutional Frictions and Interactions}} \\
\addlinespace

FX volatility           & $-1.254^{***}$ & $-1.028$      & $-.951^{**}$  & $-.922^{**}$ \\
                        & $(.473)$       & $(.703)$      & $(.375)$      & $(.414)$     \\
\addlinespace[0.25em]

CC                      & $.041$         &               & $.031$        & $.032$        \\
                        & $(.038)$       &               & $(.027)$      & $(.028)$      \\
\addlinespace[0.25em]

ERR                     &                & $.016$        & $-.004$       & $-.004$       \\
                        &                & $(.011)$      & $(.006)$      & $(.006)$      \\
\addlinespace[0.25em]

CC $\times$ ERR         &                &               & $.046^{*}$    & $.050^{*}$    \\
                        &                &               & $(.025)$      & $(.027)$      \\
\addlinespace[0.25em]

CC $\times$ FX vol.     & $.797$         &               & $-.194$       & $-.275$       \\
                        & $(.640)$       &               & $(.821)$      & $(.816)$      \\
\addlinespace[0.25em]

ERR $\times$ FX vol.    &                & $2.252$       & $-.774$       & $-.996$       \\
                        &                & $(2.423)$     & $(1.257)$     & $(1.378)$     \\
\addlinespace[0.25em]

CC $\times$ ERR $\times$ FX vol. 
                        &                &               & $6.390^{**}$  & $7.035^{**}$  \\
                        &                &               & $(3.132)$     & $(3.300)$     \\

\midrule
\addlinespace

Num. obs.               & $123118$ & $121687$ & $121687$ & $121701$ \\
Num. groups: currency   & $78$        & $78$        & $78$        & $78$        \\
Num. groups: week       & $313$       & $313$       & $313$       & $313$       \\

R$^2$ (within)          & $.364$      & $.362$      & $.370$      & $.393$      \\

\bottomrule

\end{tabular}

\begin{tablenotes}[flushleft]
\footnotesize
\item \textit{Notes:} This table reports panel regression estimates of BTC P2P premiums (Columns 1-3) and currency-specific spreads (Column 4) based on \cref{eq:hetero_inst}. \textit{CC} is constructed following \cref{eq:cc} using a threshold of $\delta = 0.7$. A robustness check using $\delta = 0.5$ is reported in \Cref{Appendix:robust}. All regressions include country and time fixed effects. Within R$^2$ is computed after removing fixed effects. Standard errors are clustered at the currency level and reported in parentheses.
\item $^{***}p<0.01$, $^{**}p<0.05$, $^{*}p<0.1$.
\end{tablenotes}

\end{threeparttable}
\end{table}

%% file: RegressionTables/remittance_weekly.tex
\begin{table}[htbp]
\centering
\begin{threeparttable}
\caption{Determinants of P2P Premiums: Substitution Effects under Financial Constraints}
\label{Tab:remittance}

\renewcommand*{\arraystretch}{1.2}
\setlength{\tabcolsep}{5pt}

\begin{tabular}{lccc}
\toprule
 & Currency-specific spread & \multicolumn{2}{c}{P2P premium} \\
\cmidrule(lr){2-2}\cmidrule(lr){3-4}
 & (1) & (2) & (3) \\
\midrule

\addlinespace
\multicolumn{4}{l}{\textit{Panel A: Market Conditions}} \\
\addlinespace

Prem$_{t-1}$            & $.578^{***}$  & $.627^{***}$  & $.630^{***}$  \\
                        & $(.035)$      & $(.031)$      & $(.031)$      \\
\addlinespace[0.25em]

P2P trading volume      & $-.007^{***}$ & $-.007^{***}$ & $-.007^{***}$ \\
                        & $(.002)$      & $(.002)$      & $(.002)$      \\
\addlinespace[0.25em]

BTC return              & $.025^{***}$  & $.020^{***}$  & $.019^{***}$  \\
                        & $(.005)$      & $(.005)$      & $(.005)$      \\
\addlinespace[0.25em]

BTC hourly volatility   & $.049^{***}$  & $-.065^{***}$ & $-.074^{***}$ \\
                        & $(.015)$      & $(.015)$      & $(.014)$      \\

\addlinespace[0.6em]
\multicolumn{4}{l}{\textit{Panel B: Remittance Costs and Constraints}} \\
\addlinespace

FX volatility           & $.036$        & $-.082$       & $-.001$       \\
                        & $(.094)$      & $(.086)$      & $(.097)$      \\
\addlinespace[0.25em]

RC                      & $-.121$       & $-.101$       & $-.101$       \\
                        & $(.082)$      & $(.075)$      & $(.076)$      \\
\addlinespace[0.25em]

Constrained $\times$ RC & $.917^{**}$   & $.808^{**}$   & $.810^{**}$   \\
                        & $(.359)$      & $(.323)$      & $(.323)$      \\
\addlinespace[0.25em]

Depr                    &               &               & $-.252^{***}$ \\
                        &               &               & $(.044)$      \\

\midrule
\addlinespace

Num. obs.               & $12289$ & $12289$ & $12257$ \\
Num. groups: currency   & $74$       & $74$       & $74$       \\
Num. groups: month      & $62$       & $62$       & $62$       \\

R$^2$ (within)          & $.453$     & $.471$     & $.474$     \\

\bottomrule

\end{tabular}

\begin{tablenotes}[flushleft]
\footnotesize
\item \textit{Notes:} This table reports panel regression estimates examining the effects of remittance costs and institutional constraints on currency-specific spreads (Column 1) and BTC P2P premiums (Columns 2 and 3). The data are aggregated at the currency-month level. \textit{RC} denotes remittance costs. \textit{Constrained} is an indicator for currencies subject to both capital controls and pegged exchange rate regimes. All regressions include country and time fixed effects. Within R$^2$ is computed after removing fixed effects. Standard errors are clustered at the currency level and reported in parentheses.
\item $^{***}p<0.01$, $^{**}p<0.05$, $^{*}p<0.1$.
\end{tablenotes}

\end{threeparttable}
\end{table}

%% file: RegressionTables/amount.tex
\begin{table}[!tbp]
\centering
\begin{threeparttable}
\caption{Determinants of P2P Trading Volume}
\label{Tab:amount}

\renewcommand*{\arraystretch}{1.2}
\setlength{\tabcolsep}{5pt}

\begin{tabular}{lccc}
\toprule
 & \multicolumn{3}{c}{Volume} \\
\cmidrule(lr){2-4}
 & (1) & (2) & (3) \\
\midrule

P2P trading volume$_{t-1}$ & $.465^{***}$  & $.460^{***}$  & $.452^{***}$  \\
                           & $(.027)$      & $(.029)$      & $(.028)$      \\
\addlinespace[0.25em]

BTC return                 & $.495^{***}$  & $.470^{***}$  & $.392^{***}$  \\
                           & $(.073)$      & $(.076)$      & $(.071)$      \\
\addlinespace[0.25em]

BTC hourly volatility     & $3.932^{***}$ & $3.812^{***}$ & $1.759^{***}$ \\
                           & $(.249)$      & $(.239)$      & $(.205)$      \\
\addlinespace[0.25em]

FX volatility              & $-6.336$      & $-4.275$      & $-6.369$      \\
                           & $(3.817)$     & $(5.022)$     & $(3.958)$     \\
\addlinespace[0.25em]

Constrained $\times$ FX volatility 
                           &               & $9.748$       &               \\
                           &               & $(19.941)$    &               \\
\addlinespace[0.25em]

Median confirmation time   &               &               & $-.011$       \\
                           &               &               & $(.019)$      \\
\addlinespace[0.25em]

On-chain cost (USD)        &               &               & $-.665^{***}$ \\
                           &               &               & $(.071)$      \\
\addlinespace[0.25em]

On-chain transactions      &               &               & $1.347^{***}$ \\
                           &               &               & $(.107)$      \\

\midrule
\addlinespace

Num. obs.                 & $134965$ & $121730$ & $134965$ \\
Num. groups: currency     & $80$        & $78$        & $80$        \\
Num. groups: week         & $313$       & $313$       & $313$       \\

R$^2$ (within)            & $.221$      & $.215$      & $.245$      \\

\bottomrule

\end{tabular}

\begin{tablenotes}[flushleft]
\footnotesize
\item \textit{Notes:} This table reports panel regression estimates for the determinants of P2P trading volume. The dependent variable is the logarithm of daily P2P trading volume in BTC. \textit{Constrained} is defined as in \cref{eq:Constrained}. All regressions include country and time fixed effects. Within R$^2$ is computed after removing fixed effects. Standard errors are clustered at the currency level and reported in parentheses.
\item $^{***}p<0.01$, $^{**}p<0.05$, $^{*}p<0.1$.
\end{tablenotes}

\end{threeparttable}
\end{table}

%% file: Sections/Price_discovery.tex
\section{Dynamic Price Discovery: Premium and Official Depreciation}\label{sec:var}

While the results from \cref{Tab:remittance} indicate that exchange rate depreciation is reflected in LB premiums, country-level evidence (e.g., \cref{fig:Argentina,fig:Nigeria}) suggests that the premiums themselves may also anticipate future depreciation, capturing forward-looking expectations about OER movements. To examine this hypothesis and allow for dynamic interactions between the P2P premium and OER depreciation, we model their joint evolution using a two-variable panel VAR framework:
\begin{equation}
\begin{aligned}
    Depr_{c,w} &= \sum_{i=1}^{l}\zeta_i Depr_{c,w-i} + \sum_{i=1}^{l}\psi_i Prem_{c,w-i} + \mu_c + \lambda_w + \varepsilon^{Depr}_{c,w}, \\
    Prem_{c,w} &= \sum_{i=1}^{l}\chi_i Depr_{c,w-i} + \sum_{i=1}^{l}\nu_i Prem_{c,w-i} + \mu_c + \lambda_w + \varepsilon^{Prem}_{c,w}
\end{aligned}
\end{equation}
where $l$ denotes the number of lags , $\mu_c$ and $\lambda_w$ represent currency and time fixed effects, respectively, and the error terms may be contemporaneously correlated across equations. We estimate the model using a fixed-effects OLS (FE-OLS) estimator, which is well-suited for unbalanced panel datasets with a large cross-sectional and time dimension. To absorb unobserved, time-invariant currency characteristics and aggregate macroeconomic trends, we include both currency and time fixed effects. In our setting, the currency fixed effects are defined at the level of currency-period segments with stable institutional regimes, reflecting the sample construction described below. 

We construct a weekly panel of currency-level observations. To ensure sufficient time-series variation, we exclude currencies with mechanically fixed OERs. To account for structural breaks captured by the Constrained dummy, we treat regime shifts as new currency units in the panel, thereby maintaining homogeneous institutional environments within each unit. Finally, we exclude short time-series segments to ensure reliable estimation of dynamic relationships. \Cref{appendix:data_var} provides full details on the sample construction.

Before estimating the Panel VAR, we assess the time-series properties of the variables. The Im-Pesaran-Shin (IPS) panel unit root test strongly rejects the null hypothesis of a unit root ($p < 0.001$), indicating stationarity. The estimated VAR satisfies the stability condition, with all eigenvalues lying strictly inside the unit circle, ensuring that impulse response functions (IRFs) are well-defined and converge over time. Furthermore, panel Granger causality tests based on the fixed-effects specification suggest a predictive relationship from the P2P premium to subsequent OER depreciation. A Wald test provides marginal evidence that lagged premiums predict future depreciation ($p = 0.059$), while the reverse relationship is not supported. While these results should not be interpreted causally, they are consistent with a lead--lag structure in which P2P market prices contain information about future OER movements.

\begin{figure}[!tbp]
    \centering
    \includegraphics[width=0.85\textwidth]{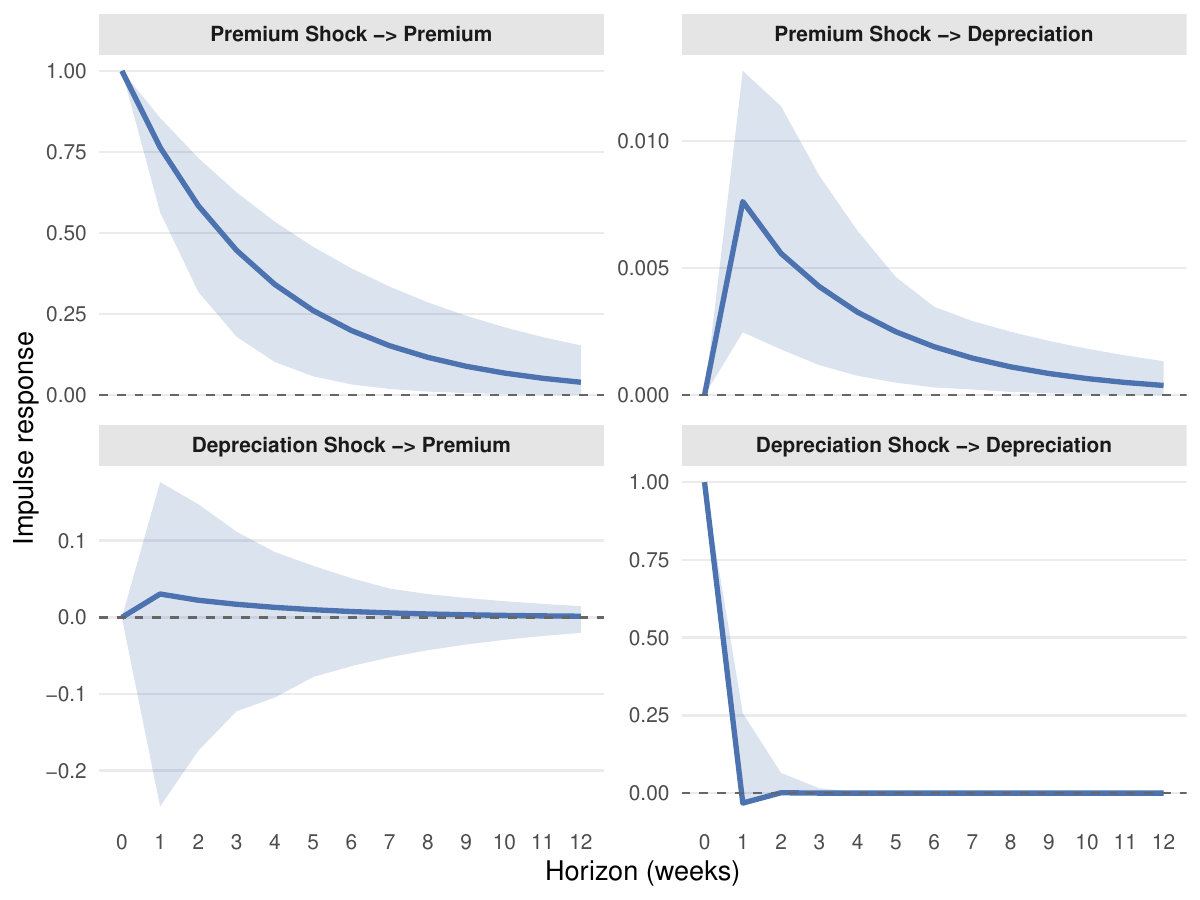}
    \caption{IRFs: P2P Premium and OER Depreciation}
    \label{fig:var}
    \begin{minipage}{\textwidth}
    \footnotesize
    \textit{Notes:} The figure reports IRFs from a panel VAR of P2P premium and OER depreciation. $l$ is set as 1 lag term. Impulse responses are shown for a one-standard-deviation shock. Shocks are identified using a Cholesky decomposition, with variables ordered as (Prem, Depr). Shaded areas represent 95\% confidence intervals, generated with bootstrap method. The horizon is measured in weeks. The VAR is estimated with currency and time fixed effects.
    \end{minipage}
\end{figure}

The impulse response analysis, reported in \cref{fig:var}, reveals that the dynamic responses differ across the two directions. A positive shock to the P2P premium leads to a statistically significant increase in future OER depreciation. The effect materializes quickly, peaks within the first week, and gradually declines while remaining positive over a 12-week horizon (Panel (b)). This pattern is consistent with the interpretation that P2P prices incorporate forward-looking information about future currency depreciation, potentially reflecting delayed adjustment in official FX markets. In contrast, shocks to OER depreciation do not generate a statistically significant response in the P2P premium (Panel (c)). Although point estimates are positive, the associated confidence intervals include zero at all horizons, indicating limited evidence of feedback from official FX markets to P2P prices.

Turning to persistence, the P2P premium exhibits strong and highly significant autocorrelation following its own shock (Panel (a)), decaying only gradually over time. This pattern indicates that the P2P premium is highly persistent, with deviations decaying only gradually over time, suggesting that the forces underlying P2P pricing adjust slowly. By contrast, depreciation shocks revert quickly to zero (Panel (d)), consistent with the view that OER movements are more tightly managed and subject to policy intervention.

As a robustness check, we explore heterogeneity by partitioning the sample into constrained and unconstrained currencies. The estimated coefficients and impulse responses display remarkably similar patterns across the two groups (see \Cref{Appendix:var_constrain}). This suggests that the ability of P2P premiums to predict future depreciation is not solely driven by extreme institutional frictions, but rather reflects a more general feature of P2P markets. Overall, the results are consistent with the interpretation that P2P prices incorporate information about future currency movements ahead of OERs.

%% file: Sections/Conclusion.tex
\section{Conclusion}\label{sec:Conclusion}

This paper studies the formation and dynamics of BTC price premiums in P2P markets across currencies with heterogeneous institutional environments. We document three key patterns. First, P2P premiums increase with FX volatility and remittance costs, particularly in constrained economies (currencies with stringent capital controls and managed exchange rate regimes), consistent with a substitution channel in which users shift toward crypto-based transactions when access to conventional cross-border payment systems is limited. Second, while P2P trading volume responds strongly to crypto-market conditions such as volatility and returns, it adjusts little to FX pressures. This asymmetry suggests that local P2P markets face supply and liquidity constraints, so that adjustment occurs primarily through prices rather than quantities. Third, P2P premiums predict subsequent OER depreciation, suggesting that these markets reflect expectations about future currency movements.

These findings point to two roles of P2P cryptocurrency markets in segmented financial systems. They provide an alternative transaction channel when access to formal cross-border payment systems is restricted, and they generate prices that reflect both current frictions and expectations about future currency movements. Because institutional and operational barriers limit arbitrage, these price gaps persist.

%% file: Sections/Appendix.tex
\clearpage
\appendix
\crefalias{section}{appendix}
\crefalias{subsection}{subappendix}
\crefalias{subsubsection}{subsubappendix}

\crefname{appendix}{Appendix}{Appendices}
\Crefname{appendix}{Appendix}{Appendices}
\crefname{subappendix}{Appendix}{Appendices}
\Crefname{subappendix}{Appendix}{Appendices}
\crefname{subsubappendix}{Appendix}{Appendices}
\Crefname{subsubappendix}{Appendix}{Appendices}

\input{Sections/Appendix_variable}

\clearpage
\input{Sections/Appendix_data}

\clearpage
\section{Robustness Checks for P2P Premium Determinants}\label{Appendix:robust}

This section provides additional robustness checks for the determinants of P2P premiums.

We assess the stability of the baseline results along two dimensions. First, we vary the definition of institutional constraints by adopting an alternative threshold (\cref{Tab:2dummy0.5}). Second, we replace the baseline remittance cost measure with alternative proxies based on sending and receiving costs (\cref{Tab:remittance_send} and \cref{Tab:remittance_receive}), which differ from the construction used in \cref{Tab:remittance} (see \cref{sec:remittance_def} for details).

Across all specifications, the main results remain qualitatively unchanged. In particular, the interaction between remittance costs and institutional constraints continues to play a central role in explaining P2P premiums, supporting the interpretation of crypto markets as a substitute channel under financial frictions.

\input{RegressionTables/2dummy0.5}
\input{RegressionTables/remittance_weekly_receive}
\input{RegressionTables/remittance_weekly_send}

\clearpage
\section{Panel VAR Results: Constrained vs. Unconstrained Currencies}\label{Appendix:var_constrain}

\begin{figure}[h]
    \centering
    \includegraphics[width=0.9\textwidth]{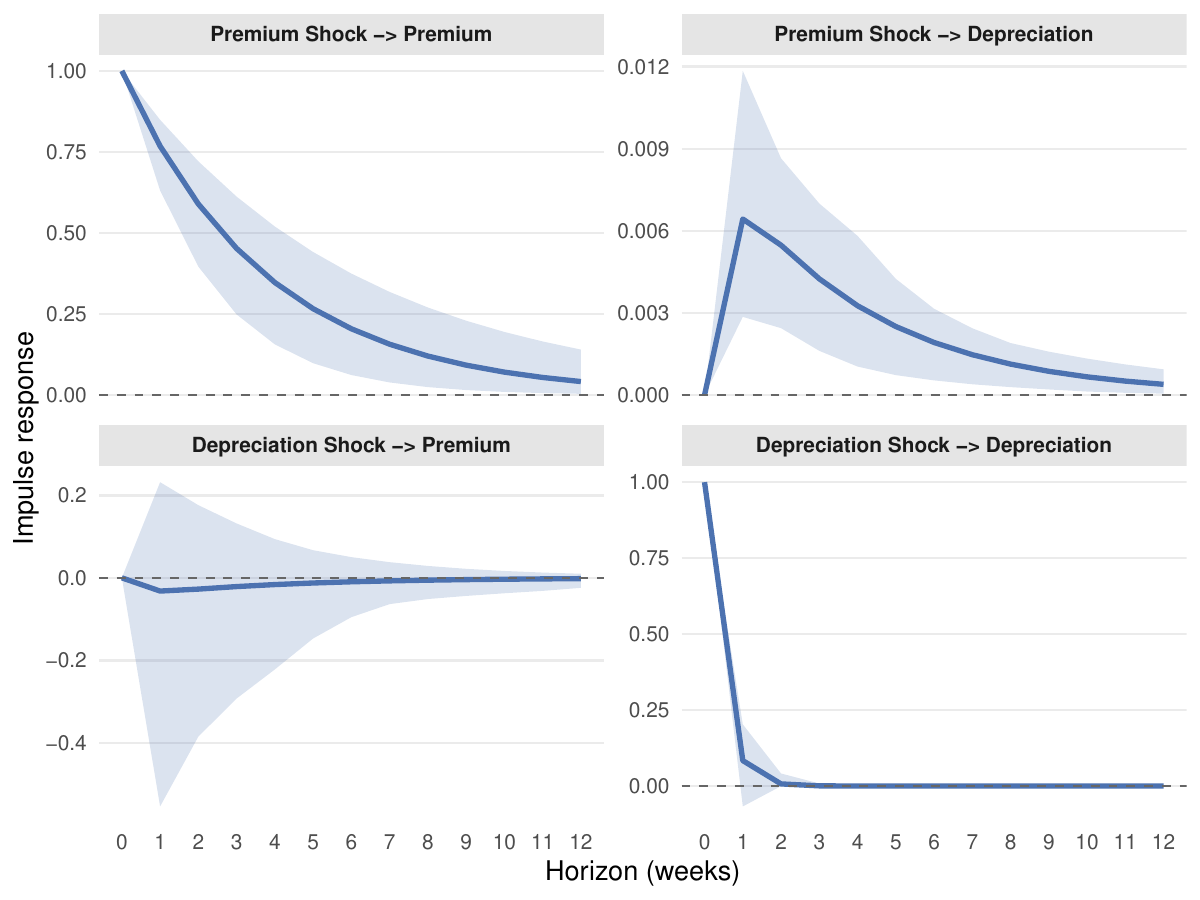}
    \caption{IRFs: Constrained Currencies}
    \label{fig:var_constrained}
    \begin{minipage}{\textwidth}
    \footnotesize
    \textit{Notes:} This figure reports IRFs for the subsample of constrained currencies. The specification is identical to that in \Cref{fig:var}. Shaded areas represent 95\% confidence intervals.
    \end{minipage}
\end{figure}

\begin{figure}[p]
    \centering
    \includegraphics[width=0.9\textwidth]{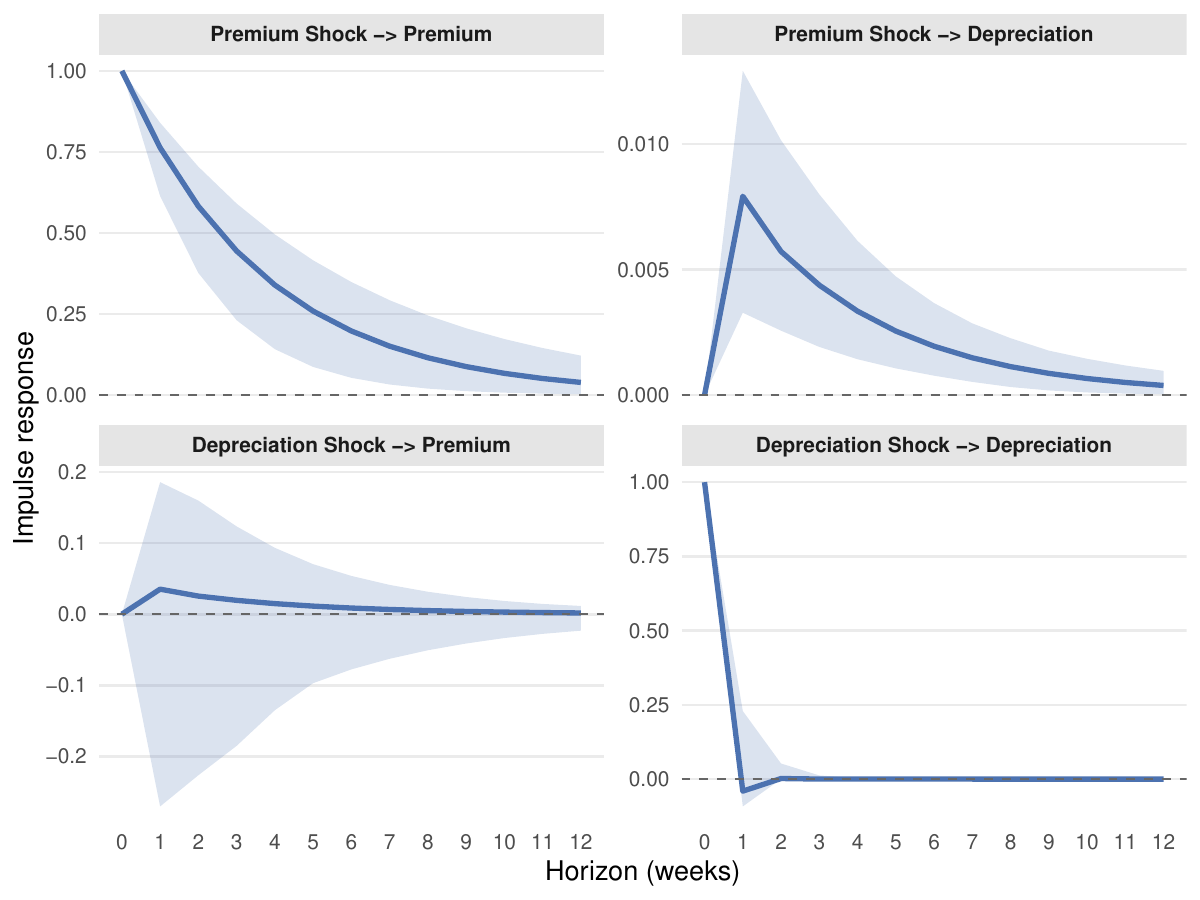}
    \caption{IRFs: Unconstrained Currencies}
    \label{fig:var_unconstrained}
    \begin{minipage}{\textwidth}
    \footnotesize
    \textit{Notes:} This figure reports IRFs for the subsample of unconstrained currencies. The specification is identical to that in \Cref{fig:var}. Shaded areas represent 95\% confidence intervals.
    \end{minipage}
\end{figure}

To assess whether the dynamic relationship between P2P premiums and OER depreciation varies across institutional environments, we re-estimate the panel VAR separately for constrained and unconstrained currencies. The results are reported in \cref{fig:var_constrained,fig:var_unconstrained}. The similarity of the estimated responses across the two subsamples suggests that the predictive content of P2P premiums is not primarily driven by either constrained or unconstrained economies, but instead reflects a more general feature of these markets.

\input{Sections/Appendix_full}

%% file: Sections/Appendix_variable.tex
\section{Variable Definitions}\label[appendix]{Appendix:variable}
This section describes the data sources and the construction of the key variables used in the analysis. We begin by providing an overview of the datasets and variables employed in the study. We then organize the variables into economically meaningful groups and describe how each set is constructed from the underlying data sources.

\subsection{Overview and Data Dictionary}\label[appendix]{Appendix:data_dict}

This subsection provides a high-level overview of all variables used in the analysis. \cref{tab:data_raw} summarizes the underlying data sources and raw measures, while \cref{tab:variables_constructed} reports the definitions and constructions of the derived variables. Additional details on temporal aggregation from daily to weekly are provided in \cref{Appendix:data_aggregation}.

\input{RegressionTables/data1}
\begin{landscape}
\thispagestyle{empty}
\begin{center}
\input{RegressionTables/data2}
\end{center}
\end{landscape}

\paragraph{Organization of variables.}
For clarity, we group variables into three categories based on their economic roles: 
\begin{itemize}
    \item P2P market dynamics and premium measures,
    \item global crypto market conditions and blockchain network frictions, and
    \item macroeconomic and institutional characteristics.
\end{itemize}

\subsection{P2P Market Dynamics and Premium Decomposition}\label[appendix]{Appendix:variable_p2p}
\subsubsection{BTC Prices on LB}
Due to the bilateral clearing of transactions on LB, we observe a wide range of prices within any given time frame. For our empirical analysis, it is necessary to aggregate this transaction-level data into a single market price to reflect the value of BTC in each currency on LB. We first generate the volume-weighted daily price:
\begin{equation}
    P^{\mathrm{P2P}}_{c,t} = \frac{\sum_{i} x_{c,i} p^{\mathrm{P2P}}_{c,i}}{\sum_{i}x_{c,i}},
    \label{eq:P2P_price}
\end{equation}
where $P^{\mathrm{P2P}}_{c,t}$ denotes the aggregated daily BTC price for currency $c$ on day $t$. For each trade $i$ executed in currency $c$ within day $t$, we use its volume $x_{c,i}$ and the corresponding BTC price $p^{\mathrm{P2P}}_{c,i}$ to calculate the volume-weighted value.

\paragraph{Data Cleaning Step.} In aggregating volume-weighted daily BTC prices on LB, we observe significant outliers that are challenging to justify, particularly given that certain trades execute at notably inflated prices despite each transaction incurring a 1\% fee. We hypothesize that these anomalies may result from transactions between acquainted users (potentially for identity verification or illicit activities) or from errors in the API logs.

Weighted average daily prices are sensitive to distortion from such extreme values, whereas median prices offer greater robustness. We identify outliers by computing the ratio of the volume-weighted average to the median daily price and examining its empirical distribution. The 1st and 99th percentiles of this ratio are 0.85 and 1.08, respectively. Values outside this range indicate that the weighted average is significantly distorted by extreme transactions; in such cases, we substitute the weighted average with the corresponding daily median.

\subsubsection{Construction of \text{P2P} Premium Components}
Based on the theoretical decomposition \cref{eq:framework_log}, the three core variables used in our baseline specifications are calculated daily for each country $c$ as follows.

The total P2P premium is calculated using the local P2P BTC price ($P^{\mathrm{P2P}}_c$), the P2P USD BTC price ($P^{\mathrm{P2P}}_{\mathrm{USD}}$), and the OER ($\mathrm{OER}_c$):
\begin{equation}
    \log(\mathrm{Prem}^{\mathrm{P2P}}_c) = \log\left( \frac{P^{\mathrm{P2P}}_c / P^{\mathrm{P2P}}_{\mathrm{USD}}}{\mathrm{OER}_c} \right)
\end{equation}

Second, the baseline P2P friction (global component) is derived from the P2P and CEX USD markets:
\begin{equation}
    \log(\text{Baseline P2P Friction}) = \log\left( \frac{P^{\mathrm{P2P}}_{\mathrm{USD}}}{P^{\mathrm{CEX}}_{\mathrm{USD}}} \right)
\end{equation}

Finally, the country-specific spread (local component) is calculated as:
\begin{equation}
    \log(\text{Country-Specific Spread}) = \log\left( \frac{P^{\mathrm{P2P}}_c / P^{\mathrm{CEX}}_{\mathrm{USD}}}{\mathrm{OER}_c} \right)
\end{equation}

\subsection{Global Crypto Market and Network Frictions}\label[appendix]{Appendix:variable_crypto}
\subsubsection{Global Crypto Market Dynamics}
In order to establish BTC price-relevant benchmarks from CEXs, we collect 5-minute tick data for BTC in USD from the Kraken cryptocurrency exchange\footnote{See \url{https://support.kraken.com/hc/en-us/articles/360047543791-Downloadable-historical-market-data-time-and-sales-}, last accessed in Mar. 2026.}. This high-frequency data allows us to precisely capture global market dynamics.

\textbf{BTC Return.} To account for global crypto market conditions, we calculate the BTC market return using log-differenced prices from the benchmark exchange. For a given frequency (daily or weekly), the return at time $t$ is calculated as:
\begin{equation}
    r^{\mathrm{CEX}}_{t} = \log\left(\frac{P^{\mathrm{CEX}}_{\mathrm{USD},t}}{P^{\mathrm{CEX}}_{\mathrm{USD},t-1}}\right), \label{eq:BTC_return}
\end{equation}
where $P^{\mathrm{CEX}}_{\mathrm{USD},t}$ represents the closing price of BTC on Kraken at time $t$.

\textbf{BTC Volatility.} We proxy BTC market volatility ($\sigma^{\mathrm{CEX}}_{w}$) using realized volatility, defined as the square root of the sum of squared log returns over a given window $w$. Based on hourly returns, realized volatility is constructed as:
\begin{equation}
    \sigma^{\mathrm{CEX}}_{w} = \sqrt{ \sum_{t \in w} \left[ \log\left(\frac{P^{\mathrm{CEX}}_{\mathrm{USD},t}}{P^{\mathrm{CEX}}_{\mathrm{USD},t-1}}\right) \right]^2 }, \label{eq:btc_vola_def}
\end{equation}
where $P^{\mathrm{CEX}}_{\mathrm{USD},t}$ denotes the BTC USD price at the interval $t$. We use hourly sampling; thus, $w$ corresponds to one day in the daily panel and one week in the weekly panel.

\subsubsection{Blockchain Network Frictions}
We collect data related to the Bitcoin network from Blockchain.com.\footnote{See \url{https://www.blockchain.com/explorer/charts}, last accessed in Mar. 2026.} Three key metrics are utilized to assess the status and efficiency of the blockchain: the median confirmation time, which represents the median duration required for a transaction to be confirmed by the BTC network (measured in minutes); the average fee in USD paid to miners for each transaction confirmed on the blockchain within a given day; and the total number of transactions confirmed on that day. The first two metrics capture the transactional risks and frictions associated with network delays and execution costs, while the latter serves as a proxy for overall network utilization and the depth of the global BTC market.

\subsection{Macroeconomic and Institutional Frictions}

\subsubsection{FX Dynamics}\label[appendix]{Appendix:fx}
\textbf{FX Depreciation.} We obtain daily FX spot rates, $\mathrm{OER}_{c,t}$, defined as the price of 1 USD in terms of currency $c$, for 80 currencies from the Thomson Reuters Refinitiv API. We utilize the midpoint of bid and ask prices at the daily close. Based on these rates, we compute the percentage depreciation of currency $c$ at time $t$ as:
\begin{equation}
    \mathrm{Depr}_{c,t} = \log\left(\frac{\mathrm{OER}_{c,t}}{\mathrm{OER}_{c,t-1}}\right). \label{eq:Depreciation}
\end{equation}

\textbf{FX Volatility.} We construct a time-varying measure of exchange rate volatility using a univariate GARCH(1,1) model estimated separately for each currency. We first compute log returns as:
\begin{equation}
    r_{c,t} := \mathrm{Depr}_{c,t}.
\end{equation}
We assume that returns follow:
\begin{equation}
    r_{c,t} = \epsilon_{c,t}, \quad \epsilon_{c,t} \sim \mathcal{N}(0, h_{c,t}),
\end{equation}
where the conditional variance evolves according to a GARCH(1,1) process:
\begin{equation}
    h_{c,t} = \omega + \alpha \epsilon_{c,t-1}^2 + \beta h_{c,t-1}.
\end{equation}

The conditional volatility is defined as:
\begin{equation}
    \sigma_{c,t} = \sqrt{h_{c,t}}.
    \label{eq:fx_vola_def}
\end{equation}
For numerical stability, returns are rescaled prior to estimation, and the resulting volatility series is subsequently transformed back to the original scale. The estimated conditional standard deviation $\sigma_{c,t}$ serves as our measure of FX volatility. All models are estimated separately for each currency using the full available sample.

For the regression analysis conducted at the weekly frequency, we aggregate daily volatility accordingly. As standard deviations are not additive, we first convert daily volatility into variance, sum within each time window (week) $w$, and then take the square root:
\begin{equation}
    \sigma_{c,w} = \left( \sum_{t \in w} \sigma_{c,t}^2 \right)^{1/2}.
    \label{eq:fx_vola_def_w}
\end{equation}
This aggregation is consistent with the properties of variance under the assumption of conditionally uncorrelated daily innovations.

\subsubsection{Institutional Frictions}\label[appendix]{Appendix:institutional}
In this study, we utilize data from the International Monetary Fund's (IMF) \textit{Annual Report on Exchange Arrangements and Exchange Restrictions} (\emph{AREAER}) to represent the degree of exchange restrictions across its 190 member countries as of 2023.\footnote{See \url{https://www.elibrary-areaer.imf.org/Pages/Home.aspx}, last accessed in March 2026.} To construct metrics for each currency, we average the metrics of countries that utilize the same currency. The country-currency pairs are detailed in \cref{Appendix:pairs}. We focus on two primary dimensions: the exchange rate regime, which indicates the extent to which a currency's exchange rate is determined by market forces, and capital controls, which regulate the flow of capital into and out of a country's economy. We consistently update our data based on the latest \emph{AREAER}, particularly for currencies that have undergone retroactive recalibrations. By tracking the effective dates of reclassification and the annual position date of each year's \emph{AREAER}, we compile this data on a daily basis.

\paragraph{Exchange Rate Regime.} Rather than relying on officially announced exchange rate arrangements, we utilize the categorization provided by the \emph{AREAER}, which focuses on de facto arrangements. These may differ from official policies but accurately reflect the actual performance and management of a currency. As outlined by \citet{habermeier2009revised}, exchange rate regimes are classified into four types---hard peg, soft peg, floating arrangements, and residual\footnote{Countries that do not fit neatly into any specific category are grouped into a residual type (e.g., nations experiencing frequent policy shifts within a short period).}---comprising 10 subcategories. This classification considers observed exchange rate behavior and interventions, such as monetary and FX policy actions. To capture the flexibility of currency $c$, we introduce an indicator variable $\mathrm{ERR}_{c,t}$ to denote whether a currency operates under a floating exchange rate regime:
\begin{equation}
    \mathrm{ERR}_{c,t} = \mathbbm{1}\left( c \text{ is not in a floating exchange rate regime} \right),
    \label{eq:ERR}
\end{equation}

\paragraph{Capital Controls.} In examining capital controls, the \emph{AREAER} offers detailed legislative information across ten asset classes, addressing inflow and outflow controls separately.\footnote{Less than 2.8\% of observations were marked as missing for the relevant data range (2017 to 2022), per the \emph{AREAER}.} To synthesize this data, we analyze each asset class $i$ and its $m_{i}$ corresponding subcategories, differentiating between controls impacting the inflow and outflow of assets. We define an indicator for capital controls on subcategory $j$ of asset $i$ at time $t$ as follows:
\begin{equation}
    a^{ijk}_{t} = \mathbbm{1}\left( \text{a capital control is in place for asset } (i,j) \right).
\end{equation}
where $k \in \{\text{inflow}, \text{outflow}\}$. By weighting all ten asset classes equally, we first construct a continuous measure of capital control intensity, which ranges from 0 to 1:
\begin{equation}
    \mathrm{CC}^{\mathrm{index}}_{c,t} = \frac{1}{2} \sum_{k} \frac{1}{10} \sum_{i=1}^{10} \frac{1}{m_i} \sum_{j=1}^{m_i} a^{ijk}_{t}.
    \label{eq:cc_index}
\end{equation}

We then define a binary indicator for capital controls based on a threshold $\delta$:
\begin{equation}
    \mathrm{CC}_{c,t} = \mathbbm{1}\left( \mathrm{CC}^{\mathrm{index}}_{c,t} \ge \delta \right),
    \label{eq:cc}
\end{equation}

Finally, we construct a combined indicator variable, $\mathrm{Constrained}_{c,t}$, which identifies currencies subject to both binding capital controls and a managed exchange rate regime:
\begin{equation}
    \mathrm{Constrained}_{c,t} =
    \mathbbm{1}\left( \mathrm{ERR}_{c,t} = 1 \;\text{and}\; \mathrm{CC}_{c,t} = 1 \right).
    \label{eq:Constrained}
\end{equation} 

\subsubsection{Remittance Costs}\label{sec:remittance_def}
To quantify the costs of international capital transfers for individuals, we rely on the cost of remittances from established channels, given their comprehensive tracking and global coverage. Remittances represent one of the largest sources of capital inflows to emerging economies. According to the World Bank, remittance flows to low- and middle-income regions reached 794 billion USD worldwide in 2023, almost doubling the size of foreign direct investment \citep{Knomad2022}. Remitting funds to many developing countries is often expensive, slow, and fraught with friction, primarily due to high transfer fees in regions with underdeveloped financial institutions. This environment incentivizes individuals to seek more cost-effective methods.

As reported by \citet{World2022}, sending \$200 incurred an average global cost of 6.09\% in 2023. The regions of Sub-Saharan Africa, the Middle East and North Africa, and Europe and Central Asia remain the most costly for remittance transfers. The average costs across major remittance corridors are detailed in \cref{Tab:tab1_remittance}, where Money Transfer Operators (MTOs) dominate with an 85\% market share. Transfer speeds vary significantly across these corridors, ranging from a few minutes to over five days. The selection of these corridors often depends on the sophistication of the financial systems, the regulatory frameworks governing capital flows in each country, and local payment behaviors.

\input{RegressionTables/remittance}

We utilize the database of remittance costs monitored by Remittance Prices Worldwide (RPW) across various Remittance Service Providers (RSPs), tracking flows from 48 sending countries to 105 receiving countries.\footnote{The World Bank, Remittance Prices Worldwide, available at \url{http://remittanceprices.worldbank.org}, last accessed in Oct. 2024.} The data spans from 2008 until February 2023 (when LB ceased operations), capturing the percentage costs associated with transferring \$200 and \$500. Additional details such as sending speed, transparency, coverage of sending corridors (ranging from a few seconds to six days or more), and pick-up methods (e.g., cash pickup or bank account transfers) are also recorded. 

Guided by the average USD volume of transactions on LB, we focus on the remittance costs associated with the \$500 benchmark. For each currency, its sending cost is calculated as the average percentage cost of sending \$500 to all available destination countries. Its receiving cost is computed similarly by averaging the cost of inbound \$500 transfers from all source countries. By calculating the average of these sending and receiving costs, we derive a comprehensive quantitative proxy for the friction of international capital transfers associated with currency $c$ at time $t$, denoted as $\mathrm{RC}_{c,t}$ (expressed in percentage terms).

%% file: RegressionTables/data1.tex
\begin{table}[htbp]\centering
\begin{threeparttable}
\caption{Data Sources and Raw Measures}
\label{tab:data_raw}

\renewcommand*{\arraystretch}{1.3}
\setlength{\tabcolsep}{5.5pt}

\begin{tabular}{p{2.9cm} p{3.3cm} p{5.0cm} p{1.9cm}}
\toprule
\textbf{Dataset} & \textbf{Raw Variable} & \textbf{Description} & \textbf{Frequency} \\
\midrule

\multicolumn{4}{l}{\textit{Panel A: Market Data}} \\
\addlinespace

\multirow{2}{*}{\href{https://localbitcoins.com/}{LocalBitcoins}} 
& $p^{\text{P2P}}_{c,i}$ 
& BTC price (in currency $c$) for trade $i$ 
& Transaction-level \\
& $x_{c,i}$ 
& Trade volume (in currency $c$) for trade $i$ 
& Transaction-level \\
\addlinespace

Refinitiv FX 
& $\text{OER}_{c}$ 
& Official exchange rate (local currency per USD) 
& Daily \\

\href{https://support.kraken.com/articles/360047543791-downloadable-historical-market-data-time-and-sales-}{Kraken CEX} 
& $P^{\text{CEX}}_{\text{USD},t}$ 
& BTC/USD price from a centralized exchange 
& Intraday (5-min) \\

\addlinespace
\multicolumn{4}{l}{\textit{Panel B: Blockchain Network Conditions}} \\
\addlinespace

\multirow{3}{*}{\href{https://www.blockchain.com/explorer/charts}{Blockchain.com}} 
& Median confirmation time 
& Median transaction confirmation time (minutes) 
& Daily \\
& On-chain cost 
& Average on-chain transaction cost (USD) 
& Daily \\
& On-chain transactions 
& Number of confirmed on-chain transactions 
& Daily \\

\addlinespace
\multicolumn{4}{l}{\textit{Panel C: Institutional and Cross-Border Frictions}} \\
\addlinespace

\multirow{2}{*}{\href{https://www.elibrary-areaer.imf.org/Pages/Home.aspx}{IMF AREAER}} 
& Capital control narratives 
& Narrative information on capital controls 
& Annual \\
& Exchange rate regime narratives 
& Narrative information on exchange rate regimes 
& Annual \\
\addlinespace

\href{http://remittanceprices.worldbank.org}{World Bank RPW} 
& Remittance cost
& Cost (in \%) of sending/receiving USD 500 
& Quarterly \\

\bottomrule
\end{tabular}

\begin{tablenotes}
\footnotesize
\item \textit{Notes:} This table summarizes the raw data sources and variables used in the analysis. Panel A reports market prices and trading activity across P2P, FX, and CEXs. Panel B contains blockchain network measurements. Panel C includes institutional variables related to capital controls, exchange rate regimes, and cross-border payment frictions. 
\end{tablenotes}

\end{threeparttable}
\end{table}

%% file: RegressionTables/data2.tex
\begin{table}[htbp]\centering
\begin{threeparttable}
\caption{Definition of Constructed Variables Used in Empirical Analysis}
\label{tab:variables_constructed}

\renewcommand*{\arraystretch}{1.4}
\setlength{\tabcolsep}{5.5pt}

\begin{tabular}{p{4cm} p{2.4cm} p{6cm} p{3cm} p{4cm}}
\toprule
\textbf{Variable} & \textbf{Symbol} & \textbf{Construction / Definition} & \textbf{Unit / Interpretation} & \textbf{Transformation (Regression)} \\
\midrule

\addlinespace[0.15em]
\multicolumn{5}{l}{\textit{Panel A: Intermediate Variables}} \\
\addlinespace[0.2em]

P2P BTC price (local) 
& $P^\text{P2P}_{c,t}$ 
& Defined in \cref{eq:P2P_price}
& Price level 
& --- \\

P2P BTC price (USD) 
& $P^\text{P2P}_{\text{USD},t}$ 
& Defined in \cref{eq:P2P_price} 
& Price level
& --- \\

\midrule
\addlinespace[0.15em]
\multicolumn{5}{l}{\textit{Panel B: Variables Used in Empirical Analysis}} \\
\addlinespace[0.2em]

Total P2P premium 
& $\mathrm{Prem}^\text{P2P}_{c,t}$ 
& Constructed as in \cref{eq:premium_def}
& Ratio
& Natural log \\

Baseline P2P friction 
& --- 
& Constructed as in \cref{eq:framework}
& Ratio
& Natural log \\

Currency-specific spread 
& --- 
& Constructed as in \cref{eq:framework}
& Ratio
& Natural log \\

P2P trading volume 
& $\text{Vol}^\text{P2P}_{c,t}$ 
& Total BTC trading volume aggregated from transaction-level data 
& BTC units
& Natural log \\

Median confirmation time 
& ---
& Median transaction confirmation time
& Level (minutes)
& Natural log \\

On-chain cost 
& ---
& Average cost per transaction
& Level, (USD)
& Natural log \\

On-chain transactions 
& ---
& Number of confirmed transactions
& Level (thousands)
& Natural log \\

FX depreciation 
& $\text{Depr}_{c,t}$
& Defined in \cref{eq:Depreciation}
& Log return 
& None \\

FX volatility 
& $\sigma_{c,t}$ 
& Defined in \cref{eq:fx_vola_def,eq:fx_vola_def_w}
& Standard deviation
& $\log(1+\sigma_{c,t})$ \\

BTC return 
& $r^{\text{CEX}}_{t}$ 
& Defined in \cref{eq:BTC_return} 
& Log return 
& None \\

BTC volatility
& $\sigma^{\text{CEX}}_{w}$
& Defined in \cref{eq:btc_vola_def} 
& Standard deviation 
& $\log(1+\sigma^{\text{CEX}}_{w})$ \\

Capital control dummy 
& $\text{CC}_{c,t}$ 
& Defined in \cref{eq:cc} 
& Binary indicator 
& None \\

Exchange rate regime dummy
& $\text{ERR}_{c,t}$ 
& Defined in \cref{eq:ERR} 
& Binary indicator 
& None \\

Constrained indicator 
& $\text{Constrained}_{c,t}$ 
& Defined in \cref{eq:Constrained}
& Binary indicator
& None \\

Remittance cost 
& $\text{RC}_{c,w}$ 
& Constructed as described in \cref{sec:remittance_def}
& Percentage terms 
& $\log(1+\text{RC}_{c,w}/100)$ \\

\bottomrule
\end{tabular}

\begin{tablenotes}
\scriptsize
\item \textit{Notes:} The last column reports how each variable enters the empirical specifications. Unless otherwise noted, strictly positive variables are transformed using natural logarithms. Variables that may take values close to zero are transformed as $\log(1+x)$. Variables defined as log returns enter regressions in their original form. Aggregation from daily to weekly frequency, when applicable, is described in \cref{Appendix:data_aggregation}.
\end{tablenotes}

\end{threeparttable}
\end{table}

%% file: RegressionTables/remittance.tex
\begin{table}[!tbp]
    \centering
\begin{threeparttable}
\caption{Average Remittance Costs (USD 200) and Market Shares by Provider Type}
\label{Tab:tab1_remittance}

\renewcommand{\arraystretch}{1.15}
\setlength{\tabcolsep}{6pt}

\begin{tabular}{l c c c c}
\toprule
\textbf{Measure} & \textbf{Bank} & \textbf{MTO} & \textbf{Mobile operator} & \textbf{Post office} \\
\midrule
Cost of sending USD 200 (\%) & 10.94 & 5.28 & 2.87 & 6.78 \\
Share of remittance corridors & -- & $>$85\% & $<$1\% & -- \\
\bottomrule
\end{tabular}

\begin{tablenotes}[flushleft]
\footnotesize
\item \textit{Notes:} This table reports average remittance costs and market shares across major provider types based on \citet{World2022}. Costs refer to the average fee (in percent) of sending USD 200 internationally. MTO denotes money transfer operators. Market shares refer to the proportion of remittance corridors served by each provider type.
\end{tablenotes}

\end{threeparttable}
\end{table}

%% file: Sections/Appendix_data.tex
\section{Temporal Aggregation and VAR Construction}\label[appendix]{Appendix:data_aggregation}

This section describes how the daily variables constructed in \cref{Appendix:variable} are transformed for the empirical analysis. In particular, we detail the aggregation procedures used to construct the weekly panel dataset and outline the additional transformations required for the panel VAR specification.

\subsection{Data Transformation and Aggregation}\label[appendix]{Appendix:transformation}

While most baseline analyses are conducted at the daily frequency, we construct a weekly panel dataset for selected specifications to harmonize data frequency across sources and reduce high-frequency noise.

Daily variables are aggregated to the weekly level according to the following rules:
\begin{itemize}
    \item \textbf{Levels and Spreads:} Price-based variables (e.g., P2P premiums, baseline P2P friction, and currency-specific spreads) and blockchain variables (e.g., median confirmation time, on-chain transaction costs, and the number of transactions) are averaged within each week.
    
    \item \textbf{Flow Variables:} Trading volumes, such as P2P trading volume, are aggregated by summation over the week.
    
    \item \textbf{Returns and Volatility:} BTC returns, FX depreciation and BTC hourly volatility are computed directly within the corresponding weekly window. Weekly FX volatility is constructed differently; see \cref{eq:fx_vola_def_w}.
    
    \item \textbf{Indicator Variables:} Dummy variables (capital controls, exchange rate regime, and $Constrained$) are defined using end-of-week values.
    
    \item \textbf{Lower-Frequency Variables:} Variables observed at lower frequencies are aligned to the weekly panel. For example, remittance costs, which are reported quarterly, are applied uniformly throughout each quarter.
\end{itemize}

\subsection{Data Construction for VAR Analysis}\label[appendix]{appendix:data_var}
\paragraph{Sample Construction.}
This section describes the construction of the panel dataset used in the VAR analysis. Taking the previous weekly P2P premiums and OER depreciations, we exclude currencies whose OERs remain mechanically fixed over the sample period (2017 - 2023) since they appear no depreciation. Such series do not provide meaningful variation for identifying dynamic relationships. This includes the following currencies: AED, BOB, IRR, JOD, OMR, PAB, QAR, SAR, and TTD, as well as the base currency USD.

\paragraph{Treatment of Regime Changes.}
To account for structural breaks in exchange rate regimes and capital control policies, we allow the panel structure to vary over time. Specifically, when a currency experiences a change in its constrained status, as captured by the $Constrained$ dummy defined in \cref{eq:Constrained}, we treat the corresponding observations as belonging to a new currency unit. Formally, each currency--time segment with a stable regime is assigned a unique identifier. This ensures that fixed effects in the VAR specification capture homogeneous institutional environments within each unit, rather than averaging across structurally different regimes.

\paragraph{Minimum Segment Length.}
To ensure reliable estimation of dynamic relationships, we exclude currency segments shorter than six weeks. Short segments may arise due to frequent regime switching or missing observations in key variables, and may introduce noise into the estimation of the VAR system.

\paragraph{Final Sample.}
After applying these filters, the final sample consists of 70 currency units and 152 currency--regime segments, covering 311 weekly observations.

\paragraph{Robustness.}
Our results are robust to alternative treatments of regime changes, including specifications that do not split currencies into multiple units and alternative minimum-length thresholds.

%% file: RegressionTables/2dummy0.5.tex
\begin{table}[htbp]
\centering
\begin{threeparttable}
\caption{Determinants of P2P Premiums: Alternative Threshold for Capital Controls ($\delta = 0.5$)}
\label{Tab:2dummy0.5}

\renewcommand*{\arraystretch}{1.3}
\setlength{\tabcolsep}{5pt}

\begin{tabular}{lcccc}
\toprule
 & \multicolumn{3}{c}{P2P premium} & Currency-specific spread \\
\cmidrule(lr){2-4} \cmidrule(lr){5-5}
 & (1) & (2) & (3) & (4) \\
\midrule

\addlinespace
\multicolumn{5}{l}{\textit{Panel A: Baseline Controls}} \\
\addlinespace

Premium$_{t-1}$         & $.570^{***}$ & $.566^{***}$ & $.553^{***}$ & $.530^{***}$ \\
                        & $(.105)$     & $(.107)$     & $(.107)$     & $(.113)$     \\
\addlinespace[0.25em]

Median confirmation time & $-.009^{***}$ & $-.009^{***}$ & $-.009^{***}$ & $-.003^{**}$ \\
                        & $(.001)$      & $(.002)$      & $(.001)$      & $(.001)$     \\
\addlinespace[0.25em]

On-chain cost (USD)     & $-.038^{***}$ & $-.038^{***}$ & $-.038^{***}$ & $-.011^{**}$ \\
                        & $(.004)$      & $(.004)$      & $(.004)$      & $(.004)$     \\
\addlinespace[0.25em]

On-chain transactions   & $.063^{***}$  & $.064^{***}$  & $.065^{***}$  & $-.019^{***}$ \\
                        & $(.005)$      & $(.005)$      & $(.005)$      & $(.006)$     \\
\addlinespace[0.25em]

P2P trade amount        & $-.009^{***}$ & $-.009^{***}$ & $-.010^{***}$ & $-.012^{***}$ \\
                        & $(.002)$      & $(.001)$      & $(.001)$      & $(.001)$     \\
\addlinespace[0.25em]

BTC return              & $.041^{***}$  & $.042^{***}$  & $.042^{***}$  & $.468^{***}$ \\
                        & $(.011)$      & $(.012)$      & $(.012)$      & $(.012)$     \\

\addlinespace[0.6em]
\multicolumn{5}{l}{\textit{Panel B: Institutional Frictions and Interactions}} \\
\addlinespace

Official FX volatility  & $-1.750^{***}$ & $-1.028$      & $-1.464^{***}$ & $-1.466^{**}$ \\
                        & $(.579)$       & $(.703)$      & $(.539)$       & $(.653)$      \\
\addlinespace[0.25em]

CC                      & $.024$         &               & $.005$         & $.004$        \\
                        & $(.037)$       &               & $(.026)$       & $(.027)$      \\
\addlinespace[0.25em]

ERR                     &                & $.016$        & $-.010$        & $-.010$       \\
                        &                & $(.011)$      & $(.007)$       & $(.007)$      \\
\addlinespace[0.25em]

CC $\times$ ERR         &                &               & $.048^{**}$    & $.052^{**}$   \\
                        &                &               & $(.024)$       & $(.026)$      \\
\addlinespace[0.25em]

CC $\times$ Official FX vol. 
                        & $1.569^{***}$  &               & $.895$         & $.896$        \\
                        & $(.545)$       &               & $(.552)$       & $(.587)$      \\
\addlinespace[0.25em]

ERR $\times$ Official FX vol. 
                        &                & $2.252$       & $-.435$        & $-.629$       \\
                        &                & $(2.423)$     & $(1.563)$      & $(1.737)$     \\
\addlinespace[0.25em]

CC $\times$ ERR $\times$ Official FX vol.
                        &                &               & $4.549^{*}$    & $5.034^{*}$   \\
                        &                &               & $(2.467)$      & $(2.588)$     \\

\midrule
\addlinespace

Num. obs.               & $123118$ & $121687$ & $121687$ & $121701$ \\
Num. groups: currency   & $78$     & $78$     & $78$     & $78$     \\
Num. groups: week       & $313$    & $313$    & $313$    & $313$    \\

R$^2$ (within)          & $.363$   & $.362$   & $.367$   & $.390$   \\

\bottomrule

\end{tabular}

\begin{tablenotes}[flushleft]
\footnotesize
\item \textit{Notes:} This table replicates the baseline specification using an alternative threshold $\delta = 0.5$ to construct \textit{CC}. All other specifications are identical to \cref{Tab:2dummy0.7}. All regressions include country and time fixed effects. Within R$^2$ is computed after removing fixed effects.
\item $^{***}p<0.01$, $^{**}p<0.05$, $^{*}p<0.1$.
\end{tablenotes}

\end{threeparttable}
\end{table}

%% file: RegressionTables/remittance_weekly_receive.tex
\begin{table}[htbp]
\centering
\begin{threeparttable}
\caption{Robustness Check: Remittance Receiving Costs}
\label{Tab:remittance_receive}

\renewcommand*{\arraystretch}{1.2}
\setlength{\tabcolsep}{5pt}

\begin{tabular}{lccc}
\toprule
 & Currency-specific spread & \multicolumn{2}{c}{P2P premium} \\
\cmidrule(lr){2-2}\cmidrule(lr){3-4}
 & (1) & (2) & (3) \\
\midrule

\addlinespace
\multicolumn{4}{l}{\textit{Panel A: Market Conditions}} \\
\addlinespace

Prem$_{t-1}$            & $.615^{***}$  & $.662^{***}$  & $.666^{***}$  \\
                        & $(.047)$      & $(.040)$      & $(.040)$      \\
\addlinespace[0.25em]

P2P trading volume      & $-.007^{***}$ & $-.007^{***}$ & $-.007^{***}$ \\
                        & $(.002)$      & $(.002)$      & $(.002)$      \\
\addlinespace[0.25em]

BTC return              & $.027^{***}$  & $.023^{***}$  & $.022^{***}$  \\
                        & $(.007)$      & $(.006)$      & $(.006)$      \\
\addlinespace[0.25em]

BTC hourly volatility   & $.053^{***}$  & $-.057^{***}$ & $-.072^{***}$ \\
                        & $(.018)$      & $(.018)$      & $(.018)$      \\

\addlinespace[0.6em]
\multicolumn{4}{l}{\textit{Panel B: Remittance Costs and Constraints}} \\
\addlinespace

FX volatility           & $.011$        & $-.088$       & $-.005$       \\
                        & $(.101)$      & $(.093)$      & $(.102)$      \\
\addlinespace[0.25em]

RC (receive)            & $-.193^{**}$  & $-.164^{*}$   & $-.161^{*}$   \\
                        & $(.093)$      & $(.087)$      & $(.085)$      \\
\addlinespace[0.25em]

Constrained $\times$ RC (receive) 
                        & $.881^{**}$   & $.770^{**}$   & $.773^{**}$   \\
                        & $(.333)$      & $(.299)$      & $(.297)$      \\
\addlinespace[0.25em]

Depr                    &               &               & $-.241^{***}$ \\
                        &               &               & $(.051)$      \\

\midrule
\addlinespace

Num. obs.               & $8832$ & $8832$ & $8809$ \\
Num. groups: currency   & $62$   & $62$   & $62$   \\
Num. groups: month      & $62$   & $62$   & $62$   \\

R$^2$ (within)          & $.482$ & $.507$ & $.510$ \\

\bottomrule

\end{tabular}

\begin{tablenotes}[flushleft]
\footnotesize
\item \textit{Notes:} This table replicates the baseline specification using remittance receiving costs. Structure and controls follow \cref{Tab:remittance}. All regressions include country and time fixed effects. Within R$^2$ is computed after removing fixed effects. Standard errors are clustered at the currency level.
\item $^{***}p<0.01$, $^{**}p<0.05$, $^{*}p<0.1$.
\end{tablenotes}

\end{threeparttable}
\end{table}

%% file: RegressionTables/remittance_weekly_send.tex
\begin{table}[htbp]
\centering
\begin{threeparttable}
\caption{Robustness Check: Remittance Sending Costs}
\label{Tab:remittance_send}

\renewcommand*{\arraystretch}{1.2}
\setlength{\tabcolsep}{5pt}

\begin{tabular}{lccc}
\toprule
 & Currency-specific spread & \multicolumn{2}{c}{P2P premium} \\
\cmidrule(lr){2-2}\cmidrule(lr){3-4}
 & (1) & (2) & (3) \\
\midrule

\addlinespace
\multicolumn{4}{l}{\textit{Panel A: Market Conditions}} \\
\addlinespace

Prem$_{t-1}$            & $.570^{***}$  & $.625^{***}$  & $.628^{***}$  \\
                        & $(.043)$      & $(.037)$      & $(.037)$      \\
\addlinespace[0.25em]

P2P trading volume      & $-.010^{***}$ & $-.009^{***}$ & $-.009^{***}$ \\
                        & $(.001)$      & $(.001)$      & $(.001)$      \\
\addlinespace[0.25em]

BTC return              & $.024^{***}$  & $.017^{**}$   & $.016^{**}$   \\
                        & $(.007)$      & $(.007)$      & $(.007)$      \\
\addlinespace[0.25em]

BTC hourly volatility   & $.051^{***}$  & $-.079^{***}$ & $-.084^{***}$ \\
                        & $(.016)$      & $(.016)$      & $(.015)$      \\

\addlinespace[0.6em]
\multicolumn{4}{l}{\textit{Panel B: Remittance Costs and Constraints}} \\
\addlinespace

FX volatility           & $.036$        & $-.076$       & $-.036$       \\
                        & $(.106)$      & $(.092)$      & $(.094)$      \\
\addlinespace[0.25em]

RC (send)               & $-.009$       & $-.004$       & $-.003$       \\
                        & $(.075)$      & $(.066)$      & $(.070)$      \\
\addlinespace[0.25em]

Constrained $\times$ RC (send) 
                        & $1.097^{***}$ & $.949^{***}$  & $.944^{***}$  \\
                        & $(.357)$      & $(.325)$      & $(.326)$      \\
\addlinespace[0.25em]

Depr                    &               &               & $-.272^{***}$ \\
                        &               &               & $(.065)$      \\

\midrule
\addlinespace

Num. obs.               & $6624$ & $6624$ & $6606$ \\
Num. groups: currency   & $47$   & $47$   & $47$   \\
Num. groups: month      & $61$   & $61$   & $61$   \\

R$^2$ (within)          & $.543$ & $.541$ & $.545$ \\

\bottomrule

\end{tabular}

\begin{tablenotes}[flushleft]
\footnotesize
\item \textit{Notes:} This table replicates the baseline specification using remittance sending costs. Structure and controls follow \cref{Tab:remittance}. All regressions include country and time fixed effects. Within R$^2$ is computed after removing fixed effects. Standard errors are clustered at the currency level.
\item $^{***}p<0.01$, $^{**}p<0.05$, $^{*}p<0.1$.
\end{tablenotes}

\end{threeparttable}
\end{table}

%% file: Sections/Appendix_full.tex
\clearpage
\section{Currencies, Countries and Exchange Rates}\label[appendix]{Appendix:pairs}

The currencies and corresponding countries that have been used in this paper are the following: Swiss Francs CHF (Switzerland), United Arab Emirates Dirham AED (United Arab Emirates), Angolan Kwanza AOA (Angola), Argentine Peso ARS (Argentina), Australian Dollar AUD (Australia, Kiribati, Nauru, Tuvalu), Bosnia and Herzegovina Convertible Mark BAM (Bosnia and Herzegovina), Bangladeshi Taka BDT (Bangladesh), Bulgarian Lev BGN (Bulgaria), Bolivian Boliviano BOB (Bolivia), Brazilian Real BRL (Brazil), Belarusian Ruble BYN (Belarus), Canadian Dollar CAD (Canada), Chilean Peso CLP (Chile), Chinese Yuan CNY (China), Colombian Peso COP (Colombia), Costa Rican Colón CRC (Costa Rica), Czech Koruna CZK (Czech Republic), Danish Krone DKK (Denmark), Dominican Peso DOP (Dominican Republic), Egyptian Pound EGP (Egypt), Ethiopian Birr ETB (Ethiopia), Euro EUR (Andorra, Austria, Belgium, Cyprus, Estonia, Finland, France, Germany, Greece, Ireland, Italy, Kosovo, Latvia, Lithuania, Luxembourg, Malta, Montenegro, The Netherlands, Portugal, San Marino, Slovak Republic, Slovenia, Spain), British Pound Sterling GBP (United Kingdom), Georgian Lari GEL (Georgia), Ghanaian Cedi GHS (Ghana), Guatemalan Quetzal GTQ (Guatemala), Hong Kong Dollar HKD (Hong Kong SAR), Honduran Lempira HNL (Honduras), Croatian Kuna HRK (Croatia), Hungarian Forint HUF (Hungary), Indonesian Rupiah IDR (Indonesia), Israeli New Shekel ILS (Israel, West Bank and Gaza), Indian Rupee INR (India), Iranian Rial IRR (Iran), Icelandic Króna ISK (Iceland), Jamaican Dollar JMD (Jamaica), Jordanian Dinar JOD (Jordan), Japanese Yen JPY (Japan), Kenyan Shilling KES (Kenya), South Korean Won KRW (Korea), Kuwaiti Dinar KWD (Kuwait), Kazakhstani Tenge KZT (Kazakhstan), Sri Lankan Rupee LKR (Sri Lanka), Moroccan Dirham MAD (Morocco), Mauritian Rupee MUR (Mauritius), Malawian Kwacha MWK (Malawi), Mexican Peso MXN (Mexico), Malaysian Ringgit MYR (Malaysia), Nigerian Naira NGN (Nigeria), Norwegian Krone NOK (Norway), New Zealand Dollar NZD (New Zealand), Omani Rial OMR (Oman), Panamanian Balboa PAB (Panama), Peruvian Sol PEN (Peru), Philippine Peso PHP (Philippines), Pakistani Rupee PKR (Pakistan), Polish Zloty PLN (Poland), Paraguayan Guarani PYG (Paraguay), Qatari Riyal QAR (Qatar), Romanian Leu RON (Romania), Serbian Dinar RSD (Serbia), Russian Ruble RUB (Russia), Rwandan Franc RWF (Rwanda), Saudi Riyal SAR (Saudi Arabia), Swedish Krona SEK (Sweden), Singapore Dollar SGD (Singapore), Eswatini Lilangeni SZL (Eswatini), Thai Baht THB (Thailand), Turkish Lira TRY (Türkiye), Trinidad and Tobago Dollar TTD (Trinidad and Tobago), New Taiwan Dollar TWD (Taiwan Province of China), Tanzanian Shilling TZS (Tanzania), Ukrainian Hryvnia UAH (Ukraine), Ugandan Shilling UGX (Uganda), United States Dollar USD (Ecuador, El Salvador, Liberia, Marshall Islands, Micronesia, Palau, Panama, Puerto Rico, Somalia, Timor-Leste, United States), Uruguayan Peso UYU (Uruguay), Vietnamese Dong VND (Vietnam), Central African CFA Franc XAF (Cameroon, Central African Republic, Chad, Republic of Congo, Equatorial Guinea, Gabon), West African CFA Franc XOF (Benin, Burkina Faso, Côte d'Ivoire, Guinea-Bissau, Mali, Niger, Senegal, Togo), South African Rand ZAR (South Africa), Zambian Kwacha ZMW (Zambia).

\begin{figure}[p]
    \centering
    \includegraphics[height=0.90\textheight]{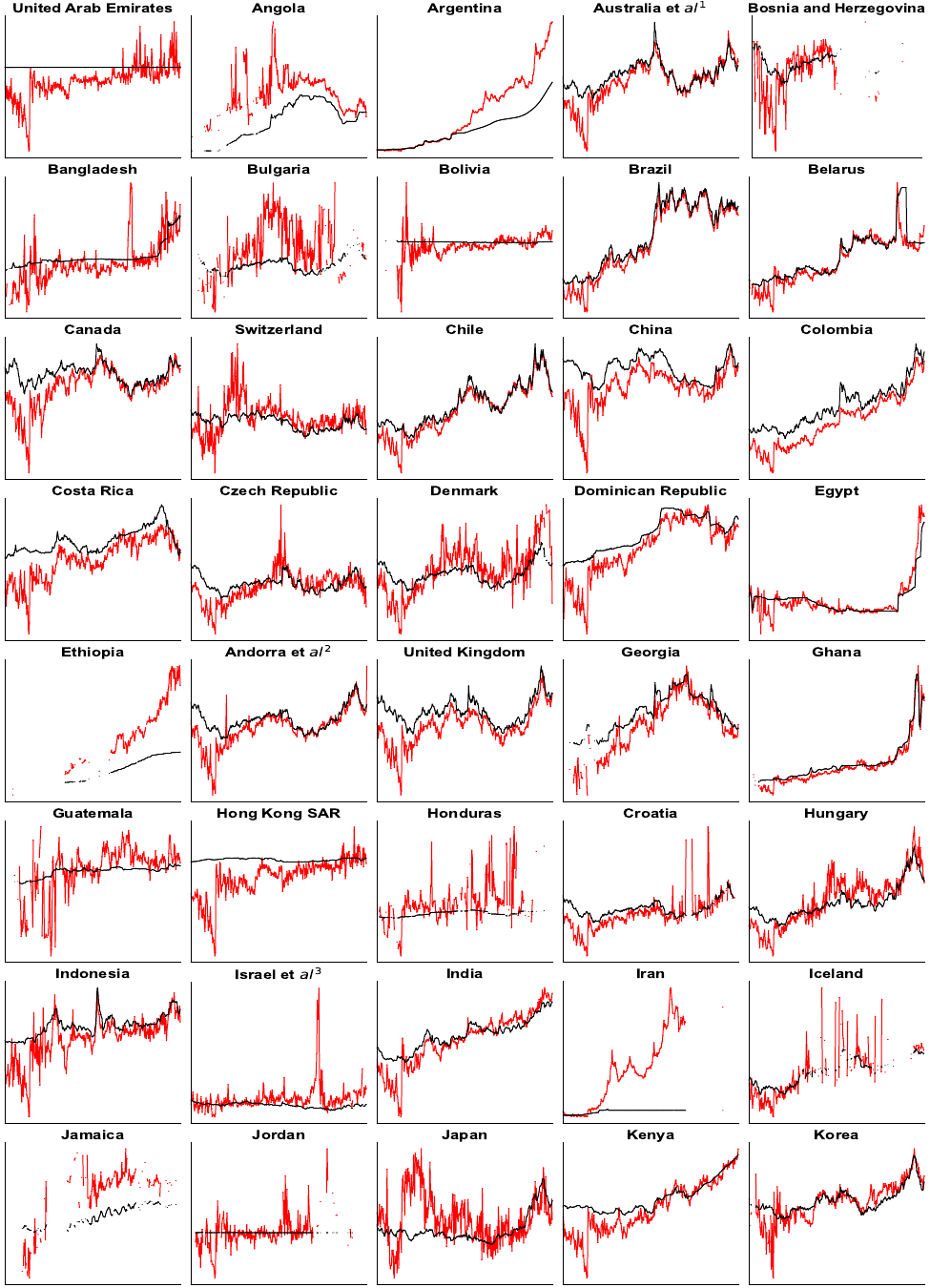}
        
    \caption{Overview of official and shadow exchange rates across countries (subset A)}
    \label{fig:OverviewPt1}
        
    \vspace{0.5em}
    \begin{minipage}{0.96\textwidth}
        \footnotesize
        \textit{Notes:} Red lines denote SERs constructed from P2P BTC prices, while black lines denote OERs. Each panel corresponds to one country and displays weekly observations over the sample period from January 1, 2017 to February 9, 2023.
    \end{minipage}
\end{figure}

\begin{figure}[p]
    \centering
    \includegraphics[height=0.90\textheight]{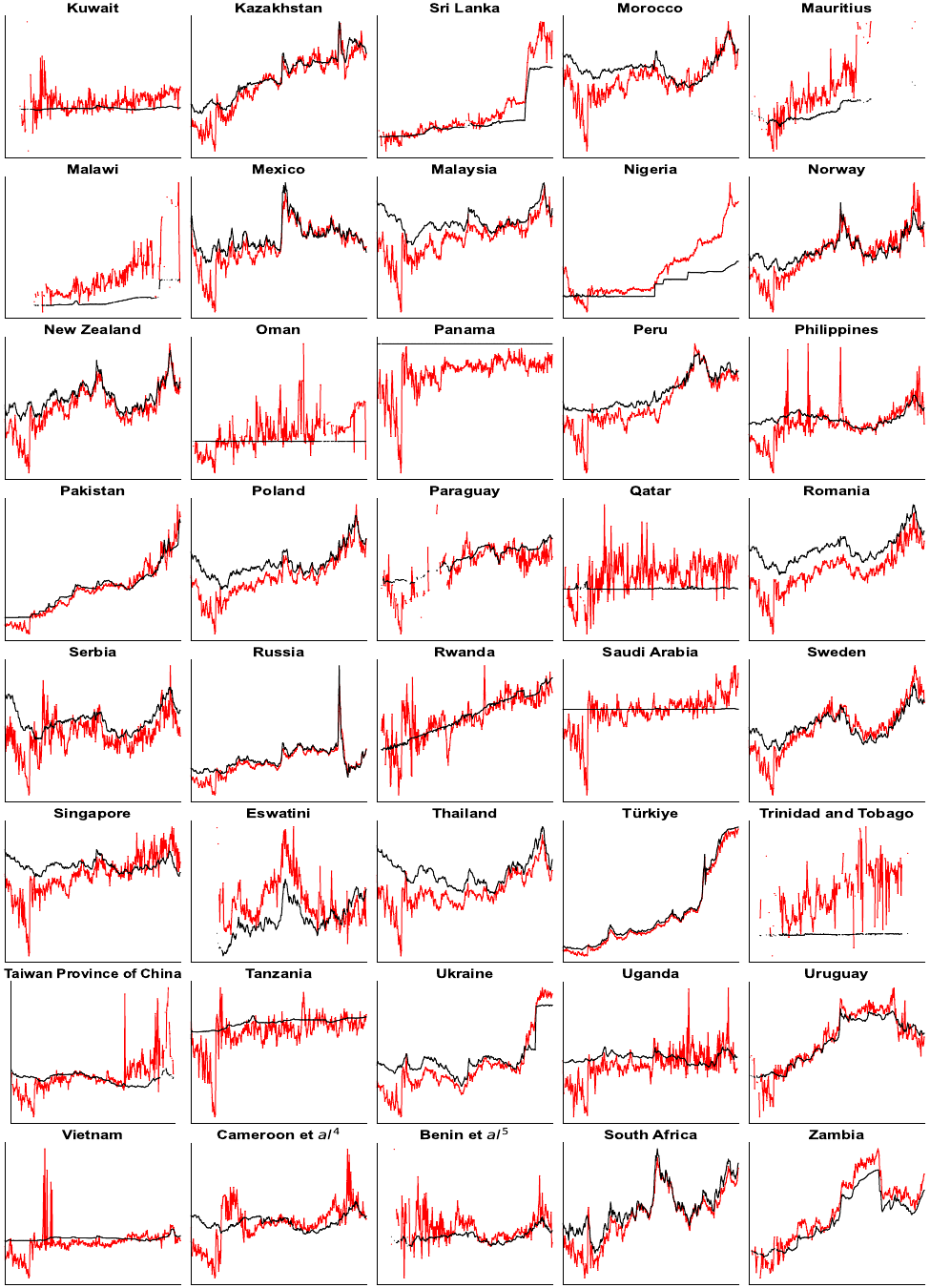}
    
    \caption{Overview of official and shadow exchange rates across countries (subset B)}
    \label{fig:OverviewPt2}
    
    \vspace{0.5em}
    \begin{minipage}{0.96\textwidth}
        \footnotesize
        \textit{Notes:} This figure reports the remaining country panels. 
    \end{minipage}

\end{figure}